\documentclass[twocolumn,showpacs,superscriptaddress,amsmath,amsfonts,amssymb]{revtex4-1}
\usepackage[T1]{fontenc}
\usepackage[latin9]{inputenc}
\usepackage{units}
\usepackage{amssymb}
\usepackage{graphicx}
\usepackage{wasysym}

\begin{document}

\title{Tracking local magnetic dynamics via high-energy charge excitations\protect\\  in a relativistic Mott insulator}

\author{N. Nembrini}

\affiliation{i-LAMP, Universit\`a Cattolica del Sacro Cuore, Brescia I-25121, Italy}

\affiliation{Department of Physics, Universit\`a degli Studi di Milano, Italy}

\author{S. Peli}

\affiliation{i-LAMP, Universit\`a Cattolica del Sacro Cuore, Brescia I-25121, Italy}

\affiliation{Department of Physics, Universit\`a degli Studi di Milano, Italy}

\author{F. Banfi}

\affiliation{i-LAMP, Universit\`a Cattolica del Sacro Cuore, Brescia I-25121, Italy}

\author{G. Ferrini}

\affiliation{i-LAMP, Universit\`a Cattolica del Sacro Cuore, Brescia I-25121, Italy}

\author{Yogesh Singh}

\affiliation{Indian Institute of Science Education and Research (IISER) Mohali, Knowledge City, Sector 81, Mohali 140306, India}

\author{P. Gegenwart}

\affiliation{Experimental Physics VI, Center for Electronic Correlations and Magnetism,
University of Augsburg, 86159 Augsburg, Germany}

\author{R. Comin}

\affiliation{Department of Physics and Astronomy, University of British Columbia, Vancouver, BC V6T 1Z1, Canada}
\affiliation{Department of Physics, Massachusetts Institute of Technology, Cambridge, Massachusetts 02139, USA}

\author{K. Foyevtsova}

\affiliation{Department of Physics and Astronomy, University of British Columbia,
Vancouver, BC V6T 1Z1, Canada}

\affiliation{Quantum Matter Institute, University of British Columbia, Vancouver,
BC V6T 1Z4, Canada}

\author{A. Damascelli}

\affiliation{Department of Physics and Astronomy, University of British Columbia,
Vancouver, BC V6T 1Z1, Canada}

\affiliation{Quantum Matter Institute, University of British Columbia, Vancouver,
BC V6T 1Z4, Canada}

\author{A. Avella}

\affiliation{Dipartimento di Fisica "E.R. Caianiello", Universit\`a degli Studi di Salerno, I-84084 Fisciano (SA), Italy}

\affiliation{CNR-SPIN, UoS di Salerno, Via Giovanni Paolo II 132, I-84084 Fisciano
(SA), Italy}

\affiliation{Unit\`a CNISM di Salerno, Universit\`a degli Studi di Salerno, I-84084
Fisciano (SA), Italy}

\author{C. Giannetti}

\affiliation{i-LAMP, Universit\`a Cattolica del Sacro Cuore, Brescia I-25121, Italy}
\begin{abstract}
We use time- and energy-resolved optical spectroscopy to investigate
the coupling of electron-hole excitations to the magnetic environment
in the relativistic Mott insulator Na$_{2}$IrO$_{3}$. We show that,
on the picosecond timescale, the photoinjected electron-hole pairs
delocalize on the hexagons of the Ir lattice via the formation of quasi-molecular
orbital (QMO) excitations and the exchange of energy with the short-range-ordered zig-zag magnetic background. 
The possibility of mapping the magnetic dynamics, which is characterized by typical frequencies in the THz range,
onto high-energy (1-2 eV) charge excitations provides a new platform to investigate, and possibly control, the dynamics of magnetic interactions in correlated materials with strong spin-orbit coupling, even in the presence of complex magnetic phases.
\end{abstract}
\maketitle


Addressing the way local charge excitations delocalize and interact with the  magnetic environment is a urgent task in the study of correlated materials. The possible impacts range from the clarification of the microscopic origin of the spin correlations that give rise to exotic magnetic states in transition-metal oxides \cite{Balents_10,Pesin2010,Tokiwa2014,Chun_15,Dean_16} to the possibility of photo-stimulating novel functionalities in materials in which the charge and magnetic excitations are strongly intertwined \cite{Kirilyuk2010,Kubacka2014,Giannetti_16}.
The family of $5d$ transition-metal oxides offers a particularly interesting playground to achieve these goals. In these materials, the on-site Coulomb repulsion $U$ is reduced to $\approx\unit[2]{eV}$ by the increased average radius
of the orbitals and the spin-orbit coupling (SOC) interaction is pushed
up to $\lambda_{SOC}\approx\unit[0.7]{eV}$ by the large mass of the
metal atoms \cite{Kim_08,Jackeli_09,Shitade_09}. As a consequence of the interplay of $U$, $\lambda_{SOC}$ and the delocalization driven by the hopping, neither a local Mott-like picture nor a delocalized orbital scenario
is fully appropriate to capture the rich phenomenology of these materials:
from Mott insulating states in which the local moments have also orbital
character \cite{Kim_08,Comin_12} to exotic spin-liquid phases \cite{Balents_10}
driven by bond-directional magnetic interactions \cite{Chun_15}.

Here, we focus on the honeycomb relativistic Mott insulator Na$_2$IrO$_3$ and adopt a non-equilibrium approach \cite{Giannetti_16}. We investigate how electron-hole excitations, photoinjected across the Mott gap \cite{Alpichshev_15,Hinton_15}, delocalize on the honeycomb lattice and release energy to the local magnetic background.  
Our experiment shows that the delocalization of local photostimulated charges is mediated by the excitations of specific QMOs. The perfect symmetry matching between the relevant QMOs and the zig-zag magnetic order allows the energy transfer to the magnetic background on the picosecond timescale. Exploiting the SOC-driven intertwining of magnetism and high-binding-energy electronic states, we show that the melting of the short-range zig-zag order is mapped onto the modification of the optical properties in the near infrared region. This finding provides a new platform to track the magnetic dynamics in 5$d$ transition metal oxides, as well as in many other relativistic correlated materials.

In the honeycomb iridate Na$_{2}$IrO$_{3}$, the crystal-field-split Ir-$t_{2g}$ orbitals
host 5 electrons \cite{Felner_02,Kobayashi_03,Singh_10}. In the
localized picture \cite{Kim_08,Chaloupka_10,Chaloupka_13}, SOC splits
the $t_{2g}$ manifold into completely filled $J_{eff}$=3/2
levels and one half-filled $J_{eff}$=1/2 state. The $J_{eff}$=1/2 level is further split by the on-site Coulomb repulsion $U$ into an empty upper-Hubbard-band (UHB) and an occupied lower-Hubbard-band (LHB), thus leading
to a relativistic Mott insulator. This scenario provides
a good explanation for both the measured insulating gap of $\approx\unit[340]{meV}$
\cite{Comin_12} and for the complex magnetic properties of the material,
which exhibits strong magnetic correlations below $\Theta_{corr}\approx\unit[120]{K}$
and a long-range-ordered zig-zag phase below $T_{\mathrm{N}}\approx\unit[15]{K}$
\cite{Singh_10,Ye_12,Choi_12}. The large frustration parameter $\Theta_{corr}/T_{\mathrm{N}}\approx8$
\cite{Singh_10} and the evidence of bond-directional magnetic interactions described
by the Kitaev model \cite{Chun_15} suggest the tendency to form a low-temperature
spin-liquid phase out of which the zig-zag order emerges.

On the other hand, the large oxygen-mediated hopping term ($\approx\unit[250]{meV}$)
between neighboring Ir atoms drives a very effective electronic delocalization
over the Ir hexagons. As a consequence, quasi-molecular orbitals (QMOs),
built from linear combinations of Ir-$t_{2g}$ Wannier functions,
emerge as the natural basis for describing the electronic properties
of Na$_{2}$IrO$_3$ \cite{Mazin_12,Foyevtsova_13}. It has been recently
demonstrated that the concept of QMOs is the key to rationalize the
outcome of \textit{ab-initio} band-structure calculations \cite{Mazin_12,Foyevtsova_13}
and accounts for the manifold of structures observed in the $\unit[0.3-2]{eV}$
energy range via photoemission and optical spectroscopy \cite{Comin_12,Li_15}.

\begin{figure}[t]
\noindent \begin{centering}
\includegraphics[width=0.48\textwidth]{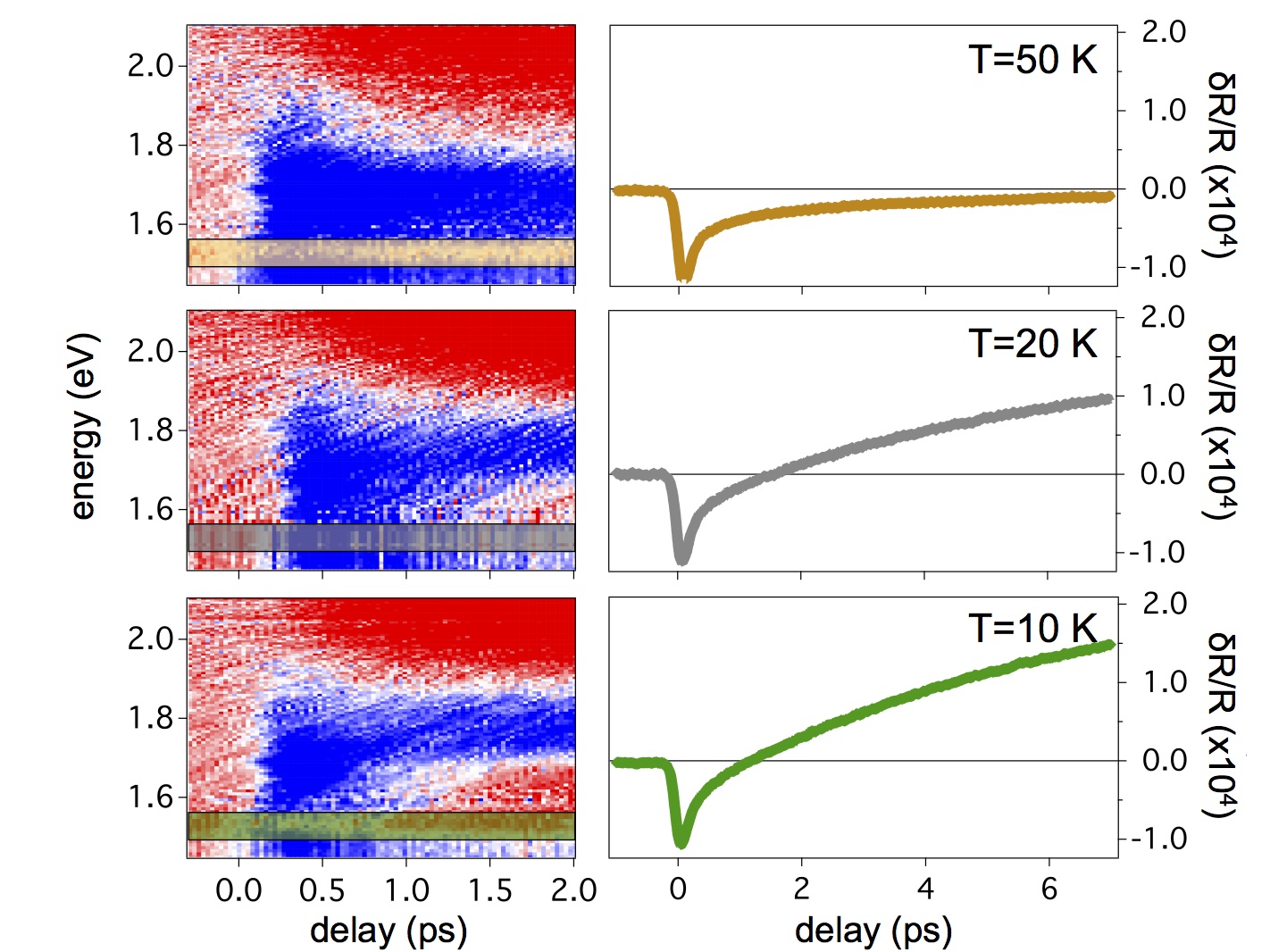}
\par\end{centering}

\caption{(Color online) Left: energy- and time- resolved reflectivity measurements
on Na$_{2}$IrO at different temperatures. The color scale
indicates positive (negative) $\delta R(\omega,t)/R$ in
red (blue). The pump fluence was varied between 10 and $\unit[40] {\mu J/cm^2}$.
Right: time traces, i.e., constant-energy cuts, extracted from $\delta R(\omega,t)/R$ for $t=\unit[2]{ps}$
at $T=50$, $20$ and $\unit[10]{K}$.\label{fig_data}}
\end{figure}

\begin{figure}[t]
\noindent \centering{}\includegraphics[width=0.48\textwidth]{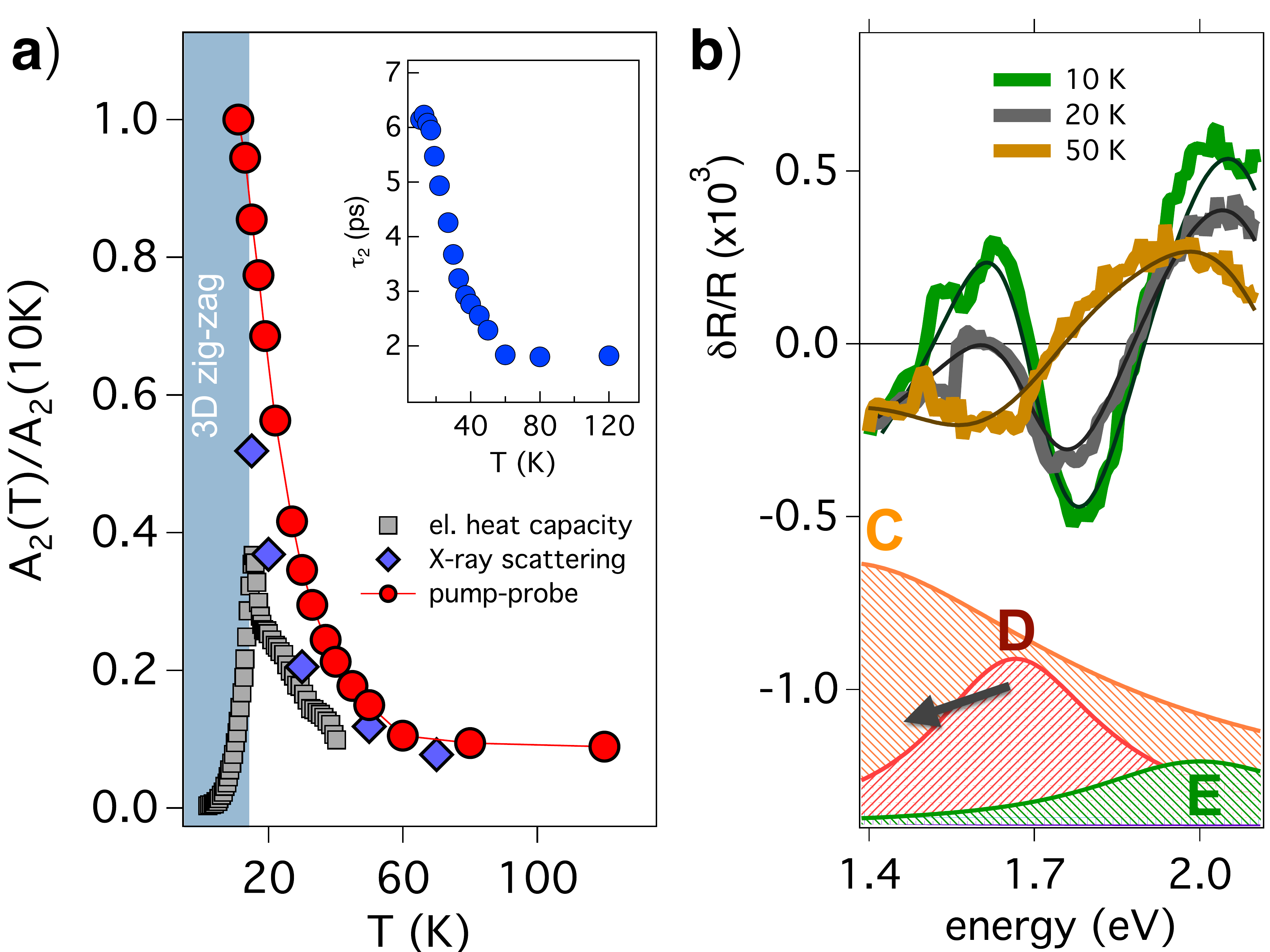}
\caption{(Color online) (a) Temperature dependence of the amplitude (red dots)
of the slow exponential function, $A{}_{2}$, renormalized to the
value at $T=\unit[10]{K}$. The values were extracted by the exponential fitting of the  $\delta R(\omega,t)/R$ time-traces for $\hbar\omega$=1.5 eV and by single-color time-resolved reflectivity measurements with fluence of $\unit[0.5-10] {\mu J/cm^2}$. The pump-probe data are compared to the
magnetic heat capacity (grey squares, taken from Ref. \citenum{Singh_10}), i.e. $\Delta C$=$C(T)$-$C_{lat}(T)$ where $C(T)$ is the total heat capacity and    and $C_{lat}(T)$ the lattice heat capacity; and to the total integral of the $H$ scan of the magnetic peak at \textbf{Q}=(0 1), as measured by resonant inelastic X-ray scattering in Ref. \citenum{Chun_15}
(blue diamonds, taken from figure S3). Both the heat capacity and the intensity of the scattering signal at the magnetic peak are reported in arbitrary units, to allow direct comparison with time-resolved data on the same graph. The blue area indicates the region of long-range 3D zig-zag order below $T_N$. The inset reports the temperature
dependence of the time constant, $\tau_{2}$, of the slow exponential
function. (b) Energy-traces extracted from $\delta R(\omega,t)/R$
for $t=\unit[2]{ps}$. The black lines represent the fit to the data
of a differential dielectric function in which only the oscillator
corresponding to the optical transition D has been modified. The fitting
procedure returns a redshift of $\approx\unit[10]{meV}$ for the oscillator
D, at a pump fluence of $\approx\unit[10] {\mu J/cm^2}$.\label{fig_analysis}}
\end{figure}

Broadband pump-probe optical spectroscopy is performed on high-quality Na$_2$IrO$_3$ crystals\cite{Singh_10}. While the pump pulse ($\hbar \omega_p$=1.55 eV photon energy) is used to photoinject local electron-hole excitations across the Mott gap, the broadband probe snaps the dynamics of the optical transitions in the range 1$-$2 eV, that corresponds to large binding-energy QMOs.
The time- and energy- resolved reflectivity variation ($\delta R(\omega,t)$/$R$)
measurements have been performed by exploiting the supercontinuum
white light produced by a photonic crystal fiber seeded by a cavity-dumped
($\unit[543]{kHz}$ rep. rate) Ti:sapphire oscillator (for details
see Refs.~\citenum{Giannetti_11} and \citenum{Cilento_10}). This configuration
allowed us to perform measurements in the low-excitation regime and avoid the local heating of the sample\cite{Supplementary}. In Fig.~\ref{fig_data}
(left panels), we report the two-dimensional $\delta R(\omega,t)$/$R$
plots taken at three different temperatures.
Following the pump excitation at $t=0$, the $\delta R(\omega,t)$/$R$
signal is characterized by a temperature-independent response, which
turns from positive (red) to negative (blue) at $\hbar\omega\approx\unit[1.8]{eV}$.
This response can be attributed to a thermomodulation effect \cite{Sun1994}
that results in the broadening of the interband transitions involving energy scales $>2$ eV.
\begin{figure*}[t]
\noindent \centering{}\includegraphics[width=1\textwidth]{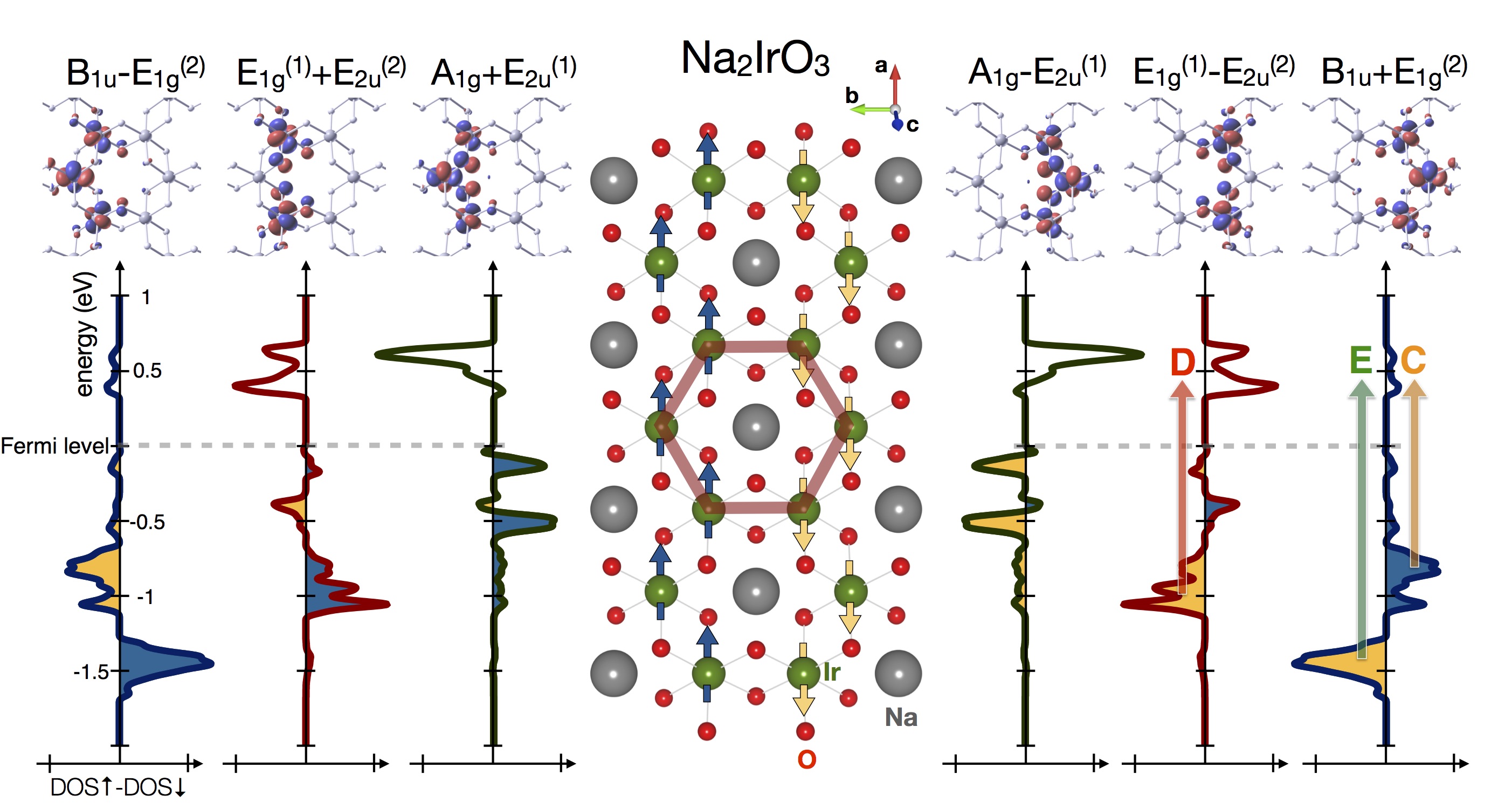}
\caption{(Color online) QMO description of the Na$_{2}$IrO$_{3}$ bandstructure.
The \textit{ab-initio} total spin-resolved DOS is projected onto suitable
combinations of the QMOs. The six linear combinations showed in the
figure reproduce both the spatial (see the wavefunctions reported
in the top panel) and the magnetic polarization (see the net polarizations,
defined as the differences between the QMO-projected spin-up and spin-down
DOSs, reported in the bottom panel) of the zig-zag ordered ground
state. The onsite Coulomb repulsion is accounted for by a $U_{eff}=\unit[2.4]{eV}$
correction that generates the experimental insulating gap \cite{Comin_12,Li_15}.
The Na$_{2}$IrO$_{3}$ honeycomb lattice structure is reported in
the central panel. The magnetic zig-zag ordered phase is indicated
by the blue (spin-up) and yellow (spin-down) arrows. The correspondence between the optical transitions C, D, E reported in Fig. \ref{fig_analysis} and the QMOs is highlighted by the colored arrows. \label{fig_DFT}}
\end{figure*}
More interestingly, a new positive component appears a few picoseconds
after the excitation in a narrow spectral region ($\unit[1.4-1.7]{eV}$)
when $T_{\mathrm{N}}$ is approached. The dynamics of this additional
component is clearly evidenced by the right panels of Fig.~\ref{fig_data},
which report the time-traces extracted from the $\delta R(\omega,t)/R$
signal at $\hbar\omega\approx\unit[1.55]{eV}$. On the picosecond timescale,
the dynamics of the $\delta R(t)/R$ signal is characterized by a
very slow build-up time of a positive component, whose amplitude increases
at low $T$. In order to systematically extract the amplitude and
timescale of this component, we performed a simple double-exponential
fit of the function $A_{1}\exp\left(-\nicefrac{t}{\tau_{1}}\right)+A_{2}\exp\left(-\nicefrac{t}{\tau_{2}}\right)$
to the data. While both $A_{1}$ and $\tau_{1}$ are found to be temperature
independent\cite{Supplementary} in the entire
explored temperature range ($\unit[10-120]{K}$), the amplitude
and timescale of the second component dramatically increase at low
temperatures, as shown in Fig.~\ref{fig_analysis}a (red dots). To rule out possible artifacts related to impulsive heating, that could be particularly relevant at low temperatures, very low-fluence (0.1-10 $\mu$J/cm$^2$) single-color ($\hbar\omega_p$=1.55 eV) time-resolved reflectivity measurements were performed\cite{Supplementary}. The values of $A_{2}$ and $\tau_{2}$ are found to be constant up to fluences of 10 $\mu$J/cm$^2$, thus demonstrating that the slowing down of the second dynamics is a genuine physical effect related to the progressive onset of magnetic correlations on approaching $T_N$. 

These results are in agreement with single-color measurements reported in Refs. \citenum{Alpichshev_15} and \citenum{Hinton_15}, which suggested the following two-step picture: i) on the $\tau_{1}\approx\unit[200]{fs}$ timescale, the high-energy electron-hole excitations created by the 1.55 eV pump pulse relax through electron-electron interactions and the coupling with optical phonons, thus leading to the fast accumulation of doublons (i.e. doubly occupied Ir sites) and holons (i.e. unoccupied Ir sites) at the bottom (top) of the UHB (LHB); ii) the interaction
with the locally ordered magnetic environment, characterized by the combination of Kitaev and Heisenberg-type correlations, drives the binding of the holons and doublons into local excitons, whose binding energy grows with their distance \cite{Alpichshev_15}. 
As a consequence of the strong dynamical constraints, the delocalization of the holon-doublon pairs is expected to be strongly delayed and to be eventually effective only on the picosecond timescale \cite{Alpichshev_15}. 

While the dynamics of the local excitons can be rationalized within the $J_{eff}$ model \cite{Alpichshev_15}, its relationship with the modification of the high-energy optical properties remained obscure so far. 
The frequency-resolution of the data reported in this work allows us to make a fundamental step beyond single-color measurements and to address both the origin of the observed signal at $\approx$1.6 eV and the way the local excitons delocalize on the Ir hexagons and release their excess energy to the zig-zag magnetic background. In particular, the results presented in Fig. \ref{fig_data} and \ref{fig_analysis}a are characterized by two important features. 
First, the $\delta R(\omega,t)$/$R$ signal in the $\unit[1.4-1.7]{eV}$ spectral range is detected at temperatures well above $T_N$ (see Fig.~\ref{fig_analysis}a), while no divergence of $A_{2}$ and $\tau_{2}$ is observed at $T_N$. This finding suggests that the relaxation dynamics of the excitons is slowed down by \textbf{short-range} zig-zag correlations \cite{Hinton_15}. This observation can be corroborated by comparing the temperature-dependence of the $A_{2}$ component to the outcomes of equilibrium techniques. In Fig.~\ref{fig_analysis}a, we report the electronic contribution to the heat capacity measured in Ref. \citenum{Singh_10}. The broad tail extending
up to $\unit[60]{K}$ was interpreted\cite{Singh_10} as the signature of the presence of short-range
magnetic correlations well above $T_{\mathrm{N}}$. More recently, the existence of zig-zag correlations at temperatures above the long-range magnetic phase transition has been directly
supported by resonant inelastic X-ray scattering measurements \cite{Chun_15}. The temperature-dependence of the magnetic diffraction peak, reported in Fig.~\ref{fig_analysis}a, directly demonstrates the persistence of zig-zag magnetic correlations, with a correlation length ($\xi$) of $\unit[1.6-1.8]{nm}$, at least up to $\approx\unit[70]{K}$. Importantly, the temperature dependence of the $A_{2}$ component in the $\delta R(t)/R$ signal almost perfectly overlaps with that of the short-range zig-zag correlations (see Fig.~\ref{fig_analysis}a). This observation strongly supports the interpretation of the slow dynamics observed in the pump-probe experiment as the time needed by the non-equilibrium distribution of electron-hole
excitations to couple to the zig-zag order on a spatial scale $\xi$, eventually
leading to its (partial) melting. 
Second, the appearance of the slow $\delta R(t)/R$ signal is confined to a narrow frequency range. This suggests that the delocalization of the doublons and the consequent melting 
of the zig-zag order selectively affect a specific high-energy QMO. Quantitatively, this can be demonstrated by a differential fitting procedure, that builds on a multi Lorentzian model of the equilibrium dielectric function\cite{Supplementary}. Both the position and the amplitude of the five oscillators used to reproduce the dielectric function perfectly map the transitions expected in the QMOs picture \cite{Li_15} (for a detailed discussion see Ref. \citenum{Supplementary}). Importantly, the spectrally-narrow positive component that appears at low temperatures is solely related to a modification of a specific oscillator (labeled D as shown in Fig. \ref{fig_analysis}), which thus unveil a direct interplay between the melting of the zig-zag order and one of the QMOs at binding energy $\geq$1 eV.

To shed light onto the relation between magnetism and QMOs, the electronic band structure of Na$_{2}$IrO$_3$ has been calculated by \textit{ab-initio}
relativistic Density Functional Theory (DFT), performed through the
linearized augmented plane wave (LAPW) method as implemented in the
full-potential code WIEN2k. As extensively discussed in several works
\cite{Mazin_12,Mazin_13,Foyevtsova_13,Li_15}, the DFT-calculated
density of states (DOS) presents five separated bands which are strongly
reminiscent of the QMOs. The six QMOs localized on a particular hexagon
can be grouped into the lowest-energy $B_{1u}$ singlet, the two doublets
$E_{1g}$ and $E_{2u}$ and the highest-energy $A_{1g}$ singlet \cite{Mazin_12}.
SOC mixes the three QMOs closer to the Fermi energy,  splits the doublets and leads to a suppression of the density of states at the Fermi level \cite{Mazin_12,Comin_12,Foyevtsova_13}. This suppression further evolves into the experimental gap when the Coulomb repulsion $U_{eff}$
is considered \cite{Comin_12,Li_15}. 

Interestingly,
the QMO representation has a strong connection with the zig-zag magnetic order emerging at low temperature. Fig.~\ref{fig_DFT} shows that suitable linear QMO combinations preserve both the spatial
arrangement and the magnetic polarization of the zig-zag ordering.
In particular, the $E_{1g}^{(1)}\pm E_{2u}^{(2)}$
combinations (dark
red curves in Fig.~\ref{fig_DFT}), which are mainly located at $\approx\unit[1]{eV}$ binding energy and account for the optical transition D,
are almost fully polarized according to the zig-zag pattern. Furthermore, they are characterized by a non-zero
overlap with both the occupied states right below the Fermi level
and the empty states at $\approx\unit[+0.3]{eV}$ (Fig. S1), thus constituting the perfect link between the low-energy dynamics
and the high-energy electronic properties. 


\begin{figure}[t]
\noindent \centering{}\includegraphics[width=0.5\textwidth]{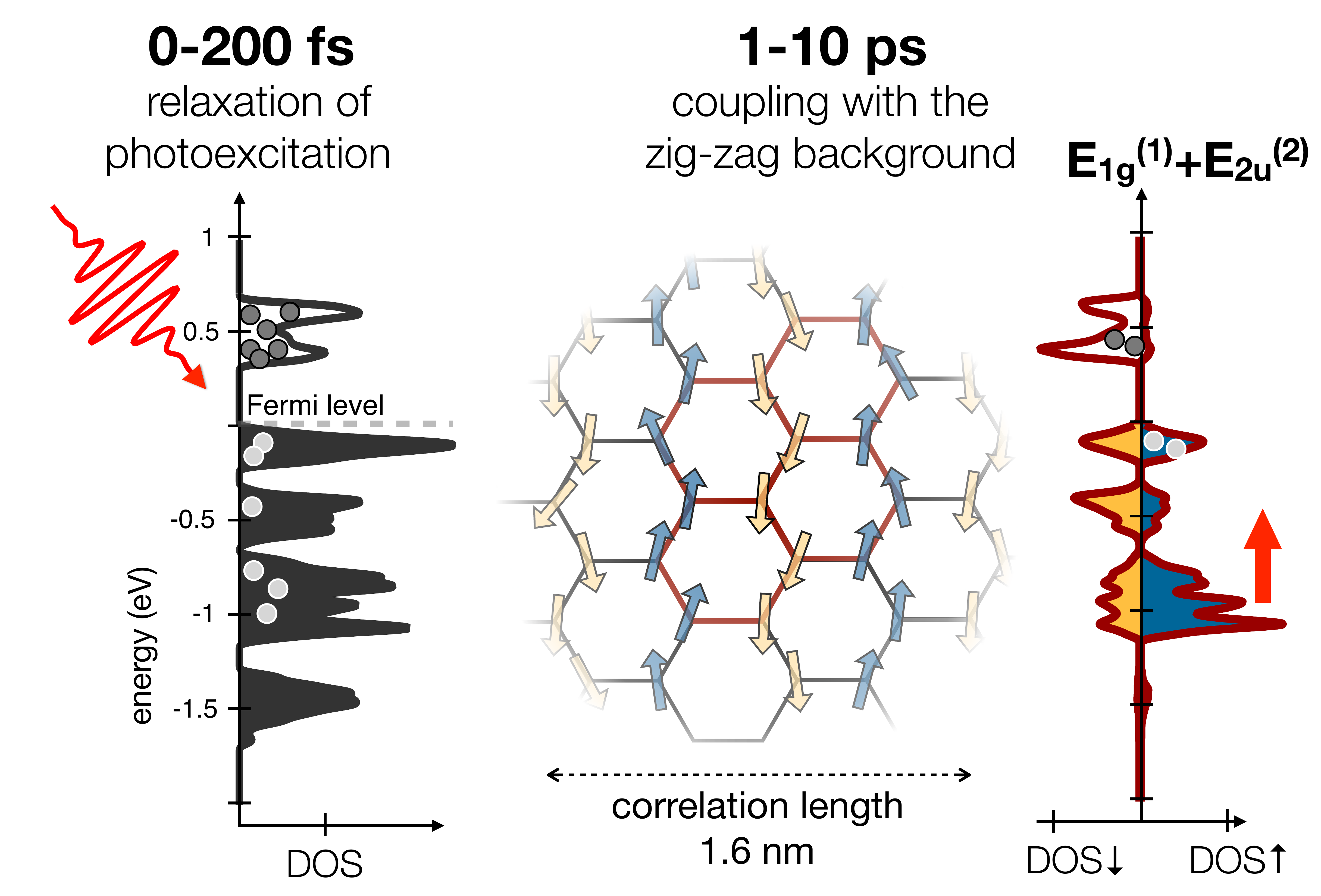}
\caption{(Color online) Cartoon of the relaxation processes after the initial
photoexcitation. The left panel reports the \textit{ab-initio} total
DOS of the occupied and unoccupied states. The right panel shows the
spin-resolved DOS of the $E_{1g}^{(1)}$+$E_{2u}^{(2)}$ QMO combination. We note that a coherence length as short as $\unit[1.6-1.8]{nm}$ coincides with
the length-scale necessary for a full tight-binding description of
the Na$_{2}$IrO$_3$ bandstructure in terms of QMOs \cite{Mazin_12}.\label{fig_summary}}
\end{figure}

The combination of non-equilibrium optics and ab-initio calculations provides a comprehensive picture of the delocalization of the holon-doublon pairs and of the energy exchange with magnetic background (see Fig.~\ref{fig_summary}). The holon-doublon pairs are initially confined by the dynamical constraints given by the topology of the zig-zag magnetic correlations \cite{Alpichshev_15}. On the picosecond timescale, the local pairs degrade via the delocalization on the Ir hexagons and form QMO-like charge excitations that are delocalized over a length scale of the order of $\unit[1.6-1.8]{nm}$. In particular, the strong $E_{1g}^{(1)}\pm E_{2u}^{(2)}$ character of the empty states at the bottom of the UHB (see Fig. \ref{fig_DFT}) allows us to identify this specific QMO combination as the channel for the energy transfer between charge excitations and the magnetic background. Considering the almost full zig-zag polarization of the $E_{1g}^{(1)}\pm E_{2u}^{(2)}$ QMOs (see Fig. \ref{fig_DFT}), we argue that the excess charge excitations in the UHB are intrinsically associated to the weakening of the zig-zag order, thus driving an effective heating of the spin background. As a feedback, the perturbation of the zig-zag magnetic order is expected to particularly affect the QMOs combinations that exhibit the same spatial pattern and spin-polarization. Specifically, the partial quench of the zig-zag magnetic correlations leads to the modification of the strongly coupled $E_{1g}^{(1)}\pm E_{2u}^{(2)}$ QMOs at $\approx\unit[-1]{eV}$ binding energy, thus allowing to map the low-energy magnetic dynamics onto the variation of the optical properties at $\unit[1.4-1.7]{eV}$. 

Our results have also a wider impact on the physics of relativistic
correlated materials. In the case of honeycomb iridates, our results demonstrate that the efficacy of the fully-localized
($J_{eff}$) and quasi-delocalized (QMO) scenarios strongly depends
on the energy scale considered: while the low-energy magnetic dynamics
is compatible with a picture of localized moments characterized
by bond-dependent (Kitaev) interactions, QMOs are the effective
building blocks of the physics at binding energies larger than $\approx\unit[1]{eV}$.
More in general, the interplay of the on-site $U$ and SOC intrinsically leads to the intertwining between the magnetic order and the high-energy electronic states in multi-orbital systems, independently of the formation of QMOs. Therefore, the possibility of mapping the demagnetization processes onto specific optical transitions in the near-infrared/visible energy range could be extended to a variety of correlated materials and frustrated magnets,
such as pyrochlore iridates and other systems that exhibit exotic
spin-liquid phases driven by Kitaev interactions. Conversely,
the resonant excitation of QMOs or other large binding-energy electronic states in relativistic correlated materials on a timescale faster than the coupling to the magnetic degrees of freedom can be used to trigger novel emergent excitations that can interact with a
\emph{cold} magnetic background and can be directly accessed
by ultrafast techniques.

We thank M. Capone, I. Elfimov, G. Jackeli, G. Khaliullin, I. Mazin for fruitful discussions. C.G., F.B. and G.F. acknowledge support from Universit\`a Cattolica del Sacro Cuore through D1, D.2.2 and D.3.1 grants.F.B. acknowledges financial support from the MIUR Futuro in ricerca 2013 Grant in the frame of the ULTRANANO Project (project number: RBFR13NEA4) and from Fondazione E.U.L.O. The work at Augsburg University is supported by the German Science Foundation through projects TRR-80 and SPP 1666. The work at UBC was supported by the Killam, A. P. Sloan, A. von Humboldt, and NSERC's Steacie Memorial Fellowships (A.D.), the Canada Research Chairs Program (A.D.), NSERC, CFI and CIFAR Quantum Materials. Y.S. acknowledges DST, India for support.

\bibliographystyle{apsrev4-1}
\bibliography{biblio_iridates}

\begin{thebibliography}{29}%
\makeatletter
\providecommand \@ifxundefined [1]{%
 \@ifx{#1\undefined}
}%
\providecommand \@ifnum [1]{%
 \ifnum #1\expandafter \@firstoftwo
 \else \expandafter \@secondoftwo
 \fi
}%
\providecommand \@ifx [1]{%
 \ifx #1\expandafter \@firstoftwo
 \else \expandafter \@secondoftwo
 \fi
}%
\providecommand \natexlab [1]{#1}%
\providecommand \enquote  [1]{``#1''}%
\providecommand \bibnamefont  [1]{#1}%
\providecommand \bibfnamefont [1]{#1}%
\providecommand \citenamefont [1]{#1}%
\providecommand \href@noop [0]{\@secondoftwo}%
\providecommand \href [0]{\begingroup \@sanitize@url \@href}%
\providecommand \@href[1]{\@@startlink{#1}\@@href}%
\providecommand \@@href[1]{\endgroup#1\@@endlink}%
\providecommand \@sanitize@url [0]{\catcode `\\12\catcode `\$12\catcode
  `\&12\catcode `\#12\catcode `\^12\catcode `\_12\catcode `\%12\relax}%
\providecommand \@@startlink[1]{}%
\providecommand \@@endlink[0]{}%
\providecommand \url  [0]{\begingroup\@sanitize@url \@url }%
\providecommand \@url [1]{\endgroup\@href {#1}{\urlprefix }}%
\providecommand \urlprefix  [0]{URL }%
\providecommand \Eprint [0]{\href }%
\providecommand \doibase [0]{http://dx.doi.org/}%
\providecommand \selectlanguage [0]{\@gobble}%
\providecommand \bibinfo  [0]{\@secondoftwo}%
\providecommand \bibfield  [0]{\@secondoftwo}%
\providecommand \translation [1]{[#1]}%
\providecommand \BibitemOpen [0]{}%
\providecommand \bibitemStop [0]{}%
\providecommand \bibitemNoStop [0]{.\EOS\space}%
\providecommand \EOS [0]{\spacefactor3000\relax}%
\providecommand \BibitemShut  [1]{\csname bibitem#1\endcsname}%
\let\auto@bib@innerbib\@empty
\bibitem [{\citenamefont {Balents}(2010)}]{Balents_10}%
  \BibitemOpen
  \bibfield  {author} {\bibinfo {author} {\bibfnamefont {L.}~\bibnamefont
  {Balents}},\ }\href@noop {} {\bibfield  {journal} {\bibinfo  {journal}
  {Nature}\ }\textbf {\bibinfo {volume} {464}},\ \bibinfo {pages} {199}
  (\bibinfo {year} {2010})}\BibitemShut {NoStop}%
\bibitem [{\citenamefont {Pesin}\ and\ \citenamefont
  {Balents}(2010)}]{Pesin2010}%
  \BibitemOpen
  \bibfield  {author} {\bibinfo {author} {\bibfnamefont {D.}~\bibnamefont
  {Pesin}}\ and\ \bibinfo {author} {\bibfnamefont {L.}~\bibnamefont
  {Balents}},\ }\href@noop {} {\bibfield  {journal} {\bibinfo  {journal}
  {Nature Physics}\ }\textbf {\bibinfo {volume} {6}},\ \bibinfo {pages} {376}
  (\bibinfo {year} {2010})}\BibitemShut {NoStop}%
\bibitem [{\citenamefont {Tokiwa}\ \emph {et~al.}(2014)\citenamefont {Tokiwa},
  \citenamefont {Ishikawa}, \citenamefont {Nakatsuji},\ and\ \citenamefont
  {Gegenwart}}]{Tokiwa2014}%
  \BibitemOpen
  \bibfield  {author} {\bibinfo {author} {\bibfnamefont {Y.}~\bibnamefont
  {Tokiwa}}, \bibinfo {author} {\bibfnamefont {J.~J.}\ \bibnamefont
  {Ishikawa}}, \bibinfo {author} {\bibfnamefont {S.}~\bibnamefont {Nakatsuji}},
  \ and\ \bibinfo {author} {\bibfnamefont {P.}~\bibnamefont {Gegenwart}},\
  }\href@noop {} {\bibfield  {journal} {\bibinfo  {journal} {Nature Materials}\
  }\textbf {\bibinfo {volume} {13}},\ \bibinfo {pages} {8} (\bibinfo {year}
  {2014})}\BibitemShut {NoStop}%
\bibitem [{\citenamefont {{Hwan Chun}}\ \emph {et~al.}(2015)\citenamefont
  {{Hwan Chun}}, \citenamefont {Kim}, \citenamefont {Kim}, \citenamefont
  {Zheng}, \citenamefont {Stoumpos}, \citenamefont {Malliakas}, \citenamefont
  {Mitchell}, \citenamefont {Mehlawat}, \citenamefont {Singh}, \citenamefont
  {Choi}, \citenamefont {Gog}, \citenamefont {Al-Zein}, \citenamefont {Sala},
  \citenamefont {Krisch}, \citenamefont {Chaloupka}, \citenamefont {Jackeli},
  \citenamefont {Khaliullin},\ and\ \citenamefont {Kim}}]{Chun_15}%
  \BibitemOpen
  \bibfield  {author} {\bibinfo {author} {\bibfnamefont {S.}~\bibnamefont
  {{Hwan Chun}}}, \bibinfo {author} {\bibfnamefont {J.-W.}\ \bibnamefont
  {Kim}}, \bibinfo {author} {\bibfnamefont {J.}~\bibnamefont {Kim}}, \bibinfo
  {author} {\bibfnamefont {H.}~\bibnamefont {Zheng}}, \bibinfo {author}
  {\bibfnamefont {C.~C.}\ \bibnamefont {Stoumpos}}, \bibinfo {author}
  {\bibfnamefont {C.~D.}\ \bibnamefont {Malliakas}}, \bibinfo {author}
  {\bibfnamefont {J.~F.}\ \bibnamefont {Mitchell}}, \bibinfo {author}
  {\bibfnamefont {K.}~\bibnamefont {Mehlawat}}, \bibinfo {author}
  {\bibfnamefont {Y.}~\bibnamefont {Singh}}, \bibinfo {author} {\bibfnamefont
  {Y.}~\bibnamefont {Choi}}, \bibinfo {author} {\bibfnamefont {T.}~\bibnamefont
  {Gog}}, \bibinfo {author} {\bibfnamefont {A.}~\bibnamefont {Al-Zein}},
  \bibinfo {author} {\bibfnamefont {M.~M.}\ \bibnamefont {Sala}}, \bibinfo
  {author} {\bibfnamefont {M.}~\bibnamefont {Krisch}}, \bibinfo {author}
  {\bibfnamefont {J.}~\bibnamefont {Chaloupka}}, \bibinfo {author}
  {\bibfnamefont {G.}~\bibnamefont {Jackeli}}, \bibinfo {author} {\bibfnamefont
  {G.}~\bibnamefont {Khaliullin}}, \ and\ \bibinfo {author} {\bibfnamefont
  {B.~J.}\ \bibnamefont {Kim}},\ }\href@noop {} {\bibfield  {journal} {\bibinfo
   {journal} {Nat. Phys.}\ }\textbf {\bibinfo {volume} {11}},\ \bibinfo {pages}
  {462} (\bibinfo {year} {2015})}\BibitemShut {NoStop}%
\bibitem [{\citenamefont {Dean}\ \emph {et~al.}(2016)\citenamefont {Dean},
  \citenamefont {Cao}, \citenamefont {Liu}, \citenamefont {Wall}, \citenamefont
  {Zhu}, \citenamefont {Mankowsky}, \citenamefont {Thampy}, \citenamefont
  {Chen}, \citenamefont {Vale}, \citenamefont {Casa}, \citenamefont {Kim},
  \citenamefont {Said}, \citenamefont {Juhas}, \citenamefont {Alonso-Mori},
  \citenamefont {Glownia}, \citenamefont {Robert}, \citenamefont {Robinson},
  \citenamefont {Sikorski}, \citenamefont {Song}, \citenamefont {Kozina},
  \citenamefont {Lemke}, \citenamefont {Patthey}, \citenamefont {Owada},
  \citenamefont {Katayama}, \citenamefont {Yabashi}, \citenamefont {Tanaka},
  \citenamefont {Togashi}, \citenamefont {Liu}, \citenamefont {{Rayan Serrao}},
  \citenamefont {Kim}, \citenamefont {Huber}, \citenamefont {Chang},
  \citenamefont {McMorrow}, \citenamefont {Forst},\ and\ \citenamefont
  {Hill}}]{Dean_16}%
  \BibitemOpen
  \bibfield  {author} {\bibinfo {author} {\bibfnamefont {M.~P.~M.}\
  \bibnamefont {Dean}}, \bibinfo {author} {\bibfnamefont {Y.}~\bibnamefont
  {Cao}}, \bibinfo {author} {\bibfnamefont {X.}~\bibnamefont {Liu}}, \bibinfo
  {author} {\bibfnamefont {S.}~\bibnamefont {Wall}}, \bibinfo {author}
  {\bibfnamefont {D.}~\bibnamefont {Zhu}}, \bibinfo {author} {\bibfnamefont
  {R.}~\bibnamefont {Mankowsky}}, \bibinfo {author} {\bibfnamefont
  {V.}~\bibnamefont {Thampy}}, \bibinfo {author} {\bibfnamefont {X.~M.}\
  \bibnamefont {Chen}}, \bibinfo {author} {\bibfnamefont {J.~G.}\ \bibnamefont
  {Vale}}, \bibinfo {author} {\bibfnamefont {D.}~\bibnamefont {Casa}}, \bibinfo
  {author} {\bibfnamefont {J.}~\bibnamefont {Kim}}, \bibinfo {author}
  {\bibfnamefont {A.~H.}\ \bibnamefont {Said}}, \bibinfo {author}
  {\bibfnamefont {P.}~\bibnamefont {Juhas}}, \bibinfo {author} {\bibfnamefont
  {R.}~\bibnamefont {Alonso-Mori}}, \bibinfo {author} {\bibfnamefont {J.~M.}\
  \bibnamefont {Glownia}}, \bibinfo {author} {\bibfnamefont {A.}~\bibnamefont
  {Robert}}, \bibinfo {author} {\bibfnamefont {J.}~\bibnamefont {Robinson}},
  \bibinfo {author} {\bibfnamefont {M.}~\bibnamefont {Sikorski}}, \bibinfo
  {author} {\bibfnamefont {S.}~\bibnamefont {Song}}, \bibinfo {author}
  {\bibfnamefont {M.}~\bibnamefont {Kozina}}, \bibinfo {author} {\bibfnamefont
  {H.}~\bibnamefont {Lemke}}, \bibinfo {author} {\bibfnamefont
  {L.}~\bibnamefont {Patthey}}, \bibinfo {author} {\bibfnamefont
  {S.}~\bibnamefont {Owada}}, \bibinfo {author} {\bibfnamefont
  {T.}~\bibnamefont {Katayama}}, \bibinfo {author} {\bibfnamefont
  {M.}~\bibnamefont {Yabashi}}, \bibinfo {author} {\bibfnamefont
  {Y.}~\bibnamefont {Tanaka}}, \bibinfo {author} {\bibfnamefont
  {T.}~\bibnamefont {Togashi}}, \bibinfo {author} {\bibfnamefont
  {J.}~\bibnamefont {Liu}}, \bibinfo {author} {\bibfnamefont {C.}~\bibnamefont
  {{Rayan Serrao}}}, \bibinfo {author} {\bibfnamefont {B.~J.}\ \bibnamefont
  {Kim}}, \bibinfo {author} {\bibfnamefont {L.}~\bibnamefont {Huber}}, \bibinfo
  {author} {\bibfnamefont {C.-L.}\ \bibnamefont {Chang}}, \bibinfo {author}
  {\bibfnamefont {D.~F.}\ \bibnamefont {McMorrow}}, \bibinfo {author}
  {\bibfnamefont {M.}~\bibnamefont {Forst}}, \ and\ \bibinfo {author}
  {\bibfnamefont {J.~P.}\ \bibnamefont {Hill}},\ }\href@noop {} {\bibfield
  {journal} {\bibinfo  {journal} {Nat. Mater.}\ }\textbf {\bibinfo {volume}
  {15}},\ \bibinfo {pages} {601} (\bibinfo {year} {2016})}\BibitemShut
  {NoStop}%
\bibitem [{\citenamefont {Kirilyuk}\ \emph {et~al.}(2010)\citenamefont
  {Kirilyuk}, \citenamefont {Kimel},\ and\ \citenamefont
  {Rasing}}]{Kirilyuk2010}%
  \BibitemOpen
  \bibfield  {author} {\bibinfo {author} {\bibfnamefont {A.}~\bibnamefont
  {Kirilyuk}}, \bibinfo {author} {\bibfnamefont {A.}~\bibnamefont {Kimel}}, \
  and\ \bibinfo {author} {\bibfnamefont {T.}~\bibnamefont {Rasing}},\
  }\href@noop {} {\bibfield  {journal} {\bibinfo  {journal} {Reviews of Modern
  Physics}\ }\textbf {\bibinfo {volume} {82}},\ \bibinfo {pages} {2731}
  (\bibinfo {year} {2010})}\BibitemShut {NoStop}%
\bibitem [{Kub()}]{Kubacka2014}%
  \BibitemOpen
  \href@noop {} {}\bibinfo {note} {Kubacka, T. et al., Science \textbf{343},
  1333 (2014)}\BibitemShut {NoStop}%
\bibitem [{\citenamefont {Giannetti}\ \emph {et~al.}(2016)\citenamefont
  {Giannetti}, \citenamefont {Capone}, \citenamefont {Fausti}, \citenamefont
  {Fabrizio}, \citenamefont {Parmigiani},\ and\ \citenamefont
  {Mihailovic}}]{Giannetti_16}%
  \BibitemOpen
  \bibfield  {author} {\bibinfo {author} {\bibfnamefont {C.}~\bibnamefont
  {Giannetti}}, \bibinfo {author} {\bibfnamefont {M.}~\bibnamefont {Capone}},
  \bibinfo {author} {\bibfnamefont {D.}~\bibnamefont {Fausti}}, \bibinfo
  {author} {\bibfnamefont {M.}~\bibnamefont {Fabrizio}}, \bibinfo {author}
  {\bibfnamefont {F.}~\bibnamefont {Parmigiani}}, \ and\ \bibinfo {author}
  {\bibfnamefont {M.}~\bibnamefont {Mihailovic}},\ }\href@noop {} {\bibfield
  {journal} {\bibinfo  {journal} {Advances in Physics}\ }\textbf {\bibinfo
  {volume} {65}},\ \bibinfo {pages} {58} (\bibinfo {year} {2016})}\BibitemShut
  {NoStop}%
\bibitem [{\citenamefont {Kim}\ \emph {et~al.}(2008)\citenamefont {Kim},
  \citenamefont {Jin}, \citenamefont {Moon}, \citenamefont {Kim}, \citenamefont
  {Park}, \citenamefont {Leem}, \citenamefont {Yu}, \citenamefont {Noh},
  \citenamefont {Kim}, \citenamefont {Oh}, \citenamefont {Park}, \citenamefont
  {Durairaj}, \citenamefont {Cao},\ and\ \citenamefont {Rotenberg}}]{Kim_08}%
  \BibitemOpen
  \bibfield  {author} {\bibinfo {author} {\bibfnamefont {B.~J.}\ \bibnamefont
  {Kim}}, \bibinfo {author} {\bibfnamefont {H.}~\bibnamefont {Jin}}, \bibinfo
  {author} {\bibfnamefont {S.~J.}\ \bibnamefont {Moon}}, \bibinfo {author}
  {\bibfnamefont {J.-Y.}\ \bibnamefont {Kim}}, \bibinfo {author} {\bibfnamefont
  {B.-G.}\ \bibnamefont {Park}}, \bibinfo {author} {\bibfnamefont {C.~S.}\
  \bibnamefont {Leem}}, \bibinfo {author} {\bibfnamefont {J.}~\bibnamefont
  {Yu}}, \bibinfo {author} {\bibfnamefont {T.~W.}\ \bibnamefont {Noh}},
  \bibinfo {author} {\bibfnamefont {C.}~\bibnamefont {Kim}}, \bibinfo {author}
  {\bibfnamefont {S.-J.}\ \bibnamefont {Oh}}, \bibinfo {author} {\bibfnamefont
  {J.-H.}\ \bibnamefont {Park}}, \bibinfo {author} {\bibfnamefont
  {V.}~\bibnamefont {Durairaj}}, \bibinfo {author} {\bibfnamefont
  {G.}~\bibnamefont {Cao}}, \ and\ \bibinfo {author} {\bibfnamefont
  {E.}~\bibnamefont {Rotenberg}},\ }\href@noop {} {\bibfield  {journal}
  {\bibinfo  {journal} {Phys. Rev. Lett.}\ }\textbf {\bibinfo {volume} {101}},\
  \bibinfo {pages} {076402} (\bibinfo {year} {2008})}\BibitemShut {NoStop}%
\bibitem [{\citenamefont {Jackeli}\ and\ \citenamefont
  {Khaliullin}(2009)}]{Jackeli_09}%
  \BibitemOpen
  \bibfield  {author} {\bibinfo {author} {\bibfnamefont {G.}~\bibnamefont
  {Jackeli}}\ and\ \bibinfo {author} {\bibfnamefont {G.}~\bibnamefont
  {Khaliullin}},\ }\href@noop {} {\bibfield  {journal} {\bibinfo  {journal}
  {Phys. Rev. Lett.}\ }\textbf {\bibinfo {volume} {102}},\ \bibinfo {pages}
  {017205} (\bibinfo {year} {2009})}\BibitemShut {NoStop}%
\bibitem [{\citenamefont {Shitade}\ \emph {et~al.}(2009)\citenamefont
  {Shitade}, \citenamefont {Katsura}, \citenamefont
  {Kune\ifmmode~\check{s}\else \v{s}\fi{}}, \citenamefont {Qi}, \citenamefont
  {Zhang},\ and\ \citenamefont {Nagaosa}}]{Shitade_09}%
  \BibitemOpen
  \bibfield  {author} {\bibinfo {author} {\bibfnamefont {A.}~\bibnamefont
  {Shitade}}, \bibinfo {author} {\bibfnamefont {H.}~\bibnamefont {Katsura}},
  \bibinfo {author} {\bibfnamefont {J.}~\bibnamefont
  {Kune\ifmmode~\check{s}\else \v{s}\fi{}}}, \bibinfo {author} {\bibfnamefont
  {X.-L.}\ \bibnamefont {Qi}}, \bibinfo {author} {\bibfnamefont {S.-C.}\
  \bibnamefont {Zhang}}, \ and\ \bibinfo {author} {\bibfnamefont
  {N.}~\bibnamefont {Nagaosa}},\ }\href@noop {} {\bibfield  {journal} {\bibinfo
   {journal} {Phys. Rev. Lett.}\ }\textbf {\bibinfo {volume} {102}},\ \bibinfo
  {pages} {256403} (\bibinfo {year} {2009})}\BibitemShut {NoStop}%
\bibitem [{\citenamefont {Comin}\ \emph {et~al.}(2012)\citenamefont {Comin},
  \citenamefont {Levy}, \citenamefont {Ludbrook}, \citenamefont {Zhu},
  \citenamefont {Veenstra}, \citenamefont {Rosen}, \citenamefont {Singh},
  \citenamefont {Gegenwart}, \citenamefont {Stricker}, \citenamefont {Hancock},
  \citenamefont {van~der Marel}, \citenamefont {Elfimov},\ and\ \citenamefont
  {Damascelli}}]{Comin_12}%
  \BibitemOpen
  \bibfield  {author} {\bibinfo {author} {\bibfnamefont {R.}~\bibnamefont
  {Comin}}, \bibinfo {author} {\bibfnamefont {G.}~\bibnamefont {Levy}},
  \bibinfo {author} {\bibfnamefont {B.}~\bibnamefont {Ludbrook}}, \bibinfo
  {author} {\bibfnamefont {Z.-H.}\ \bibnamefont {Zhu}}, \bibinfo {author}
  {\bibfnamefont {C.~N.}\ \bibnamefont {Veenstra}}, \bibinfo {author}
  {\bibfnamefont {J.~A.}\ \bibnamefont {Rosen}}, \bibinfo {author}
  {\bibfnamefont {Y.}~\bibnamefont {Singh}}, \bibinfo {author} {\bibfnamefont
  {P.}~\bibnamefont {Gegenwart}}, \bibinfo {author} {\bibfnamefont
  {D.}~\bibnamefont {Stricker}}, \bibinfo {author} {\bibfnamefont {J.~N.}\
  \bibnamefont {Hancock}}, \bibinfo {author} {\bibfnamefont {D.}~\bibnamefont
  {van~der Marel}}, \bibinfo {author} {\bibfnamefont {I.~S.}\ \bibnamefont
  {Elfimov}}, \ and\ \bibinfo {author} {\bibfnamefont {A.}~\bibnamefont
  {Damascelli}},\ }\href@noop {} {\bibfield  {journal} {\bibinfo  {journal}
  {Phys. Rev. Lett.}\ }\textbf {\bibinfo {volume} {109}},\ \bibinfo {pages}
  {266406} (\bibinfo {year} {2012})}\BibitemShut {NoStop}%
\bibitem [{\citenamefont {Alpichshev}\ \emph {et~al.}(2015)\citenamefont
  {Alpichshev}, \citenamefont {Mahmood}, \citenamefont {Cao},\ and\
  \citenamefont {Gedik}}]{Alpichshev_15}%
  \BibitemOpen
  \bibfield  {author} {\bibinfo {author} {\bibfnamefont {Z.}~\bibnamefont
  {Alpichshev}}, \bibinfo {author} {\bibfnamefont {F.}~\bibnamefont {Mahmood}},
  \bibinfo {author} {\bibfnamefont {G.}~\bibnamefont {Cao}}, \ and\ \bibinfo
  {author} {\bibfnamefont {N.}~\bibnamefont {Gedik}},\ }\href@noop {}
  {\bibfield  {journal} {\bibinfo  {journal} {Phys. Rev. Lett.}\ }\textbf
  {\bibinfo {volume} {114}},\ \bibinfo {pages} {017203} (\bibinfo {year}
  {2015})}\BibitemShut {NoStop}%
\bibitem [{\citenamefont {Hinton}\ \emph {et~al.}(2015)\citenamefont {Hinton},
  \citenamefont {Patankar}, \citenamefont {Thewalt}, \citenamefont {Ruiz},
  \citenamefont {Lopez}, \citenamefont {Breznay}, \citenamefont {Vishwanath},
  \citenamefont {Analytis}, \citenamefont {Orenstein}, \citenamefont
  {Koralek},\ and\ \citenamefont {Kimchi}}]{Hinton_15}%
  \BibitemOpen
  \bibfield  {author} {\bibinfo {author} {\bibfnamefont {J.~P.}\ \bibnamefont
  {Hinton}}, \bibinfo {author} {\bibfnamefont {S.}~\bibnamefont {Patankar}},
  \bibinfo {author} {\bibfnamefont {E.}~\bibnamefont {Thewalt}}, \bibinfo
  {author} {\bibfnamefont {A.}~\bibnamefont {Ruiz}}, \bibinfo {author}
  {\bibfnamefont {G.}~\bibnamefont {Lopez}}, \bibinfo {author} {\bibfnamefont
  {N.}~\bibnamefont {Breznay}}, \bibinfo {author} {\bibfnamefont
  {A.}~\bibnamefont {Vishwanath}}, \bibinfo {author} {\bibfnamefont
  {J.}~\bibnamefont {Analytis}}, \bibinfo {author} {\bibfnamefont
  {J.}~\bibnamefont {Orenstein}}, \bibinfo {author} {\bibfnamefont {J.~D.}\
  \bibnamefont {Koralek}}, \ and\ \bibinfo {author} {\bibfnamefont
  {I.}~\bibnamefont {Kimchi}},\ }\href@noop {} {\bibfield  {journal} {\bibinfo
  {journal} {Phys. Rev. B}\ }\textbf {\bibinfo {volume} {92}},\ \bibinfo
  {pages} {115154} (\bibinfo {year} {2015})}\BibitemShut {NoStop}%
\bibitem [{\citenamefont {Felner}\ and\ \citenamefont
  {Bradaric}(2002)}]{Felner_02}%
  \BibitemOpen
  \bibfield  {author} {\bibinfo {author} {\bibfnamefont {I.}~\bibnamefont
  {Felner}}\ and\ \bibinfo {author} {\bibfnamefont {I.}~\bibnamefont
  {Bradaric}},\ }\href {\doibase 10.1016/S0921-4526(01)01038-9} {\bibfield
  {journal} {\bibinfo  {journal} {Physica B: Condensed Matter}\ }\textbf
  {\bibinfo {volume} {311}},\ \bibinfo {pages} {195} (\bibinfo {year}
  {2002})}\BibitemShut {NoStop}%
\bibitem [{\citenamefont {Kobayashi}\ \emph {et~al.}(2003)\citenamefont
  {Kobayashi}, \citenamefont {Tabuchi}, \citenamefont {Shikano}, \citenamefont
  {Kageyama},\ and\ \citenamefont {Kanno}}]{Kobayashi_03}%
  \BibitemOpen
  \bibfield  {author} {\bibinfo {author} {\bibfnamefont {H.}~\bibnamefont
  {Kobayashi}}, \bibinfo {author} {\bibfnamefont {M.}~\bibnamefont {Tabuchi}},
  \bibinfo {author} {\bibfnamefont {M.}~\bibnamefont {Shikano}}, \bibinfo
  {author} {\bibfnamefont {H.}~\bibnamefont {Kageyama}}, \ and\ \bibinfo
  {author} {\bibfnamefont {R.}~\bibnamefont {Kanno}},\ }\href {\doibase
  10.1039/B207282C} {\bibfield  {journal} {\bibinfo  {journal} {J. Mater.
  Chem.}\ }\textbf {\bibinfo {volume} {13}},\ \bibinfo {pages} {957} (\bibinfo
  {year} {2003})}\BibitemShut {NoStop}%
\bibitem [{\citenamefont {Singh}\ and\ \citenamefont
  {Gegenwart}(2010)}]{Singh_10}%
  \BibitemOpen
  \bibfield  {author} {\bibinfo {author} {\bibfnamefont {Y.}~\bibnamefont
  {Singh}}\ and\ \bibinfo {author} {\bibfnamefont {P.}~\bibnamefont
  {Gegenwart}},\ }\href@noop {} {\bibfield  {journal} {\bibinfo  {journal}
  {Phys. Rev. B}\ }\textbf {\bibinfo {volume} {82}},\ \bibinfo {pages} {064412}
  (\bibinfo {year} {2010})}\BibitemShut {NoStop}%
\bibitem [{\citenamefont {Chaloupka}\ \emph {et~al.}(2010)\citenamefont
  {Chaloupka}, \citenamefont {Jackeli},\ and\ \citenamefont
  {Khaliullin}}]{Chaloupka_10}%
  \BibitemOpen
  \bibfield  {author} {\bibinfo {author} {\bibfnamefont {J.}~\bibnamefont
  {Chaloupka}}, \bibinfo {author} {\bibfnamefont {G.}~\bibnamefont {Jackeli}},
  \ and\ \bibinfo {author} {\bibfnamefont {G.}~\bibnamefont {Khaliullin}},\
  }\href@noop {} {\bibfield  {journal} {\bibinfo  {journal} {Phys. Rev. Lett.}\
  }\textbf {\bibinfo {volume} {105}},\ \bibinfo {pages} {027204} (\bibinfo
  {year} {2010})}\BibitemShut {NoStop}%
\bibitem [{\citenamefont {Chaloupka}\ \emph {et~al.}(2013)\citenamefont
  {Chaloupka}, \citenamefont {Jackeli},\ and\ \citenamefont
  {Khaliullin}}]{Chaloupka_13}%
  \BibitemOpen
  \bibfield  {author} {\bibinfo {author} {\bibfnamefont {J.}~\bibnamefont
  {Chaloupka}}, \bibinfo {author} {\bibfnamefont {G.}~\bibnamefont {Jackeli}},
  \ and\ \bibinfo {author} {\bibfnamefont {G.}~\bibnamefont {Khaliullin}},\
  }\href@noop {} {\bibfield  {journal} {\bibinfo  {journal} {Phys. Rev. Lett.}\
  }\textbf {\bibinfo {volume} {110}},\ \bibinfo {pages} {097204} (\bibinfo
  {year} {2013})}\BibitemShut {NoStop}%
\bibitem [{\citenamefont {Ye}\ \emph {et~al.}(2012)\citenamefont {Ye},
  \citenamefont {Chi}, \citenamefont {Cao}, \citenamefont {Chakoumakos},
  \citenamefont {Fernandez-Baca}, \citenamefont {Custelcean}, \citenamefont
  {Qi}, \citenamefont {Korneta},\ and\ \citenamefont {Cao}}]{Ye_12}%
  \BibitemOpen
  \bibfield  {author} {\bibinfo {author} {\bibfnamefont {F.}~\bibnamefont
  {Ye}}, \bibinfo {author} {\bibfnamefont {S.}~\bibnamefont {Chi}}, \bibinfo
  {author} {\bibfnamefont {H.}~\bibnamefont {Cao}}, \bibinfo {author}
  {\bibfnamefont {B.~C.}\ \bibnamefont {Chakoumakos}}, \bibinfo {author}
  {\bibfnamefont {J.~A.}\ \bibnamefont {Fernandez-Baca}}, \bibinfo {author}
  {\bibfnamefont {R.}~\bibnamefont {Custelcean}}, \bibinfo {author}
  {\bibfnamefont {T.~F.}\ \bibnamefont {Qi}}, \bibinfo {author} {\bibfnamefont
  {O.~B.}\ \bibnamefont {Korneta}}, \ and\ \bibinfo {author} {\bibfnamefont
  {G.}~\bibnamefont {Cao}},\ }\href@noop {} {\bibfield  {journal} {\bibinfo
  {journal} {Phys. Rev. B}\ }\textbf {\bibinfo {volume} {85}},\ \bibinfo
  {pages} {180403} (\bibinfo {year} {2012})}\BibitemShut {NoStop}%
\bibitem [{\citenamefont {Choi}\ \emph {et~al.}(2012)\citenamefont {Choi},
  \citenamefont {Coldea}, \citenamefont {Kolmogorov}, \citenamefont
  {Lancaster}, \citenamefont {Mazin}, \citenamefont {Blundell}, \citenamefont
  {Radaelli}, \citenamefont {Singh}, \citenamefont {Gegenwart}, \citenamefont
  {Choi}, \citenamefont {Cheong}, \citenamefont {Baker}, \citenamefont
  {Stock},\ and\ \citenamefont {Taylor}}]{Choi_12}%
  \BibitemOpen
  \bibfield  {author} {\bibinfo {author} {\bibfnamefont {S.~K.}\ \bibnamefont
  {Choi}}, \bibinfo {author} {\bibfnamefont {R.}~\bibnamefont {Coldea}},
  \bibinfo {author} {\bibfnamefont {A.~N.}\ \bibnamefont {Kolmogorov}},
  \bibinfo {author} {\bibfnamefont {T.}~\bibnamefont {Lancaster}}, \bibinfo
  {author} {\bibfnamefont {I.~I.}\ \bibnamefont {Mazin}}, \bibinfo {author}
  {\bibfnamefont {S.~J.}\ \bibnamefont {Blundell}}, \bibinfo {author}
  {\bibfnamefont {P.~G.}\ \bibnamefont {Radaelli}}, \bibinfo {author}
  {\bibfnamefont {Y.}~\bibnamefont {Singh}}, \bibinfo {author} {\bibfnamefont
  {P.}~\bibnamefont {Gegenwart}}, \bibinfo {author} {\bibfnamefont {K.~R.}\
  \bibnamefont {Choi}}, \bibinfo {author} {\bibfnamefont {S.-W.}\ \bibnamefont
  {Cheong}}, \bibinfo {author} {\bibfnamefont {P.~J.}\ \bibnamefont {Baker}},
  \bibinfo {author} {\bibfnamefont {C.}~\bibnamefont {Stock}}, \ and\ \bibinfo
  {author} {\bibfnamefont {J.}~\bibnamefont {Taylor}},\ }\href@noop {}
  {\bibfield  {journal} {\bibinfo  {journal} {Phys. Rev. Lett.}\ }\textbf
  {\bibinfo {volume} {108}},\ \bibinfo {pages} {127204} (\bibinfo {year}
  {2012})}\BibitemShut {NoStop}%
\bibitem [{\citenamefont {Mazin}\ \emph {et~al.}(2012)\citenamefont {Mazin},
  \citenamefont {Jeschke}, \citenamefont {Foyevtsova}, \citenamefont
  {Valent\'i},\ and\ \citenamefont {Khomskii}}]{Mazin_12}%
  \BibitemOpen
  \bibfield  {author} {\bibinfo {author} {\bibfnamefont {I.~I.}\ \bibnamefont
  {Mazin}}, \bibinfo {author} {\bibfnamefont {H.~O.}\ \bibnamefont {Jeschke}},
  \bibinfo {author} {\bibfnamefont {K.}~\bibnamefont {Foyevtsova}}, \bibinfo
  {author} {\bibfnamefont {R.}~\bibnamefont {Valent\'i}}, \ and\ \bibinfo
  {author} {\bibfnamefont {D.~I.}\ \bibnamefont {Khomskii}},\ }\href@noop {}
  {\bibfield  {journal} {\bibinfo  {journal} {Phys. Rev. Lett.}\ }\textbf
  {\bibinfo {volume} {109}},\ \bibinfo {pages} {197201} (\bibinfo {year}
  {2012})}\BibitemShut {NoStop}%
\bibitem [{\citenamefont {Foyevtsova}\ \emph {et~al.}(2013)\citenamefont
  {Foyevtsova}, \citenamefont {Jeschke}, \citenamefont {Mazin}, \citenamefont
  {Khomskii},\ and\ \citenamefont {Valent\'\i}}]{Foyevtsova_13}%
  \BibitemOpen
  \bibfield  {author} {\bibinfo {author} {\bibfnamefont {K.}~\bibnamefont
  {Foyevtsova}}, \bibinfo {author} {\bibfnamefont {H.~O.}\ \bibnamefont
  {Jeschke}}, \bibinfo {author} {\bibfnamefont {I.~I.}\ \bibnamefont {Mazin}},
  \bibinfo {author} {\bibfnamefont {D.~I.}\ \bibnamefont {Khomskii}}, \ and\
  \bibinfo {author} {\bibfnamefont {R.}~\bibnamefont {Valent\'\i}},\
  }\href@noop {} {\bibfield  {journal} {\bibinfo  {journal} {Phys. Rev. B}\
  }\textbf {\bibinfo {volume} {88}},\ \bibinfo {pages} {035107} (\bibinfo
  {year} {2013})}\BibitemShut {NoStop}%
\bibitem [{\citenamefont {Li}\ \emph {et~al.}(2015)\citenamefont {Li},
  \citenamefont {Foyevtsova}, \citenamefont {Jeschke},\ and\ \citenamefont
  {Valent\'i}}]{Li_15}%
  \BibitemOpen
  \bibfield  {author} {\bibinfo {author} {\bibfnamefont {Y.}~\bibnamefont
  {Li}}, \bibinfo {author} {\bibfnamefont {K.}~\bibnamefont {Foyevtsova}},
  \bibinfo {author} {\bibfnamefont {H.~O.}\ \bibnamefont {Jeschke}}, \ and\
  \bibinfo {author} {\bibfnamefont {R.}~\bibnamefont {Valent\'i}},\ }\href@noop
  {} {\bibfield  {journal} {\bibinfo  {journal} {Phys. Rev. B}\ }\textbf
  {\bibinfo {volume} {91}},\ \bibinfo {pages} {161101} (\bibinfo {year}
  {2015})}\BibitemShut {NoStop}%
\bibitem [{\citenamefont {Giannetti}\ \emph {et~al.}(2011)\citenamefont
  {Giannetti}, \citenamefont {Cilento}, \citenamefont {Conte}, \citenamefont
  {Coslovich}, \citenamefont {Ferrini}, \citenamefont {Molegraaf},
  \citenamefont {Raichle}, \citenamefont {Liang}, \citenamefont {Eisaki},
  \citenamefont {Greven}, \citenamefont {Damascelli}, \citenamefont {van~der
  Marel},\ and\ \citenamefont {Parmigiani}}]{Giannetti_11}%
  \BibitemOpen
  \bibfield  {author} {\bibinfo {author} {\bibfnamefont {C.}~\bibnamefont
  {Giannetti}}, \bibinfo {author} {\bibfnamefont {F.}~\bibnamefont {Cilento}},
  \bibinfo {author} {\bibfnamefont {S.~D.}\ \bibnamefont {Conte}}, \bibinfo
  {author} {\bibfnamefont {G.}~\bibnamefont {Coslovich}}, \bibinfo {author}
  {\bibfnamefont {G.}~\bibnamefont {Ferrini}}, \bibinfo {author} {\bibfnamefont
  {H.}~\bibnamefont {Molegraaf}}, \bibinfo {author} {\bibfnamefont
  {M.}~\bibnamefont {Raichle}}, \bibinfo {author} {\bibfnamefont
  {R.}~\bibnamefont {Liang}}, \bibinfo {author} {\bibfnamefont
  {H.}~\bibnamefont {Eisaki}}, \bibinfo {author} {\bibfnamefont
  {M.}~\bibnamefont {Greven}}, \bibinfo {author} {\bibfnamefont
  {A.}~\bibnamefont {Damascelli}}, \bibinfo {author} {\bibfnamefont
  {D.}~\bibnamefont {van~der Marel}}, \ and\ \bibinfo {author} {\bibfnamefont
  {F.}~\bibnamefont {Parmigiani}},\ }\href@noop {} {\bibfield  {journal}
  {\bibinfo  {journal} {Nat. Commun.}\ }\textbf {\bibinfo {volume} {2}},\
  \bibinfo {pages} {353} (\bibinfo {year} {2011})}\BibitemShut {NoStop}%
\bibitem [{\citenamefont {Cilento}\ \emph {et~al.}(2010)\citenamefont
  {Cilento}, \citenamefont {Giannetti}, \citenamefont {Ferrini}, \citenamefont
  {Dal~Conte}, \citenamefont {Sala}, \citenamefont {Coslovich}, \citenamefont
  {Rini}, \citenamefont {Cavalleri},\ and\ \citenamefont
  {Parmigiani}}]{Cilento_10}%
  \BibitemOpen
  \bibfield  {author} {\bibinfo {author} {\bibfnamefont {F.}~\bibnamefont
  {Cilento}}, \bibinfo {author} {\bibfnamefont {C.}~\bibnamefont {Giannetti}},
  \bibinfo {author} {\bibfnamefont {G.}~\bibnamefont {Ferrini}}, \bibinfo
  {author} {\bibfnamefont {S.}~\bibnamefont {Dal~Conte}}, \bibinfo {author}
  {\bibfnamefont {T.}~\bibnamefont {Sala}}, \bibinfo {author} {\bibfnamefont
  {G.}~\bibnamefont {Coslovich}}, \bibinfo {author} {\bibfnamefont
  {M.}~\bibnamefont {Rini}}, \bibinfo {author} {\bibfnamefont {A.}~\bibnamefont
  {Cavalleri}}, \ and\ \bibinfo {author} {\bibfnamefont {F.}~\bibnamefont
  {Parmigiani}},\ }\href@noop {} {\bibfield  {journal} {\bibinfo  {journal}
  {Applied Physics Letters}\ }\textbf {\bibinfo {volume} {96}},\ \bibinfo {eid}
  {021102} (\bibinfo {year} {2010})}\BibitemShut {NoStop}%
\bibitem [{Sup()}]{Supplementary}%
  \BibitemOpen
  \href@noop {} {}\bibinfo {note} {See Supplemental Material at [] for
  technical details on the DFT calculations, including the relation between the
  optical conductivity and the calculated band-structure, as well as additional
  details on the differential fitting and on the experiment, including the
  effect of local heating and the sub-ps dynamics.}\BibitemShut {Stop}%
\bibitem [{\citenamefont {Sun}\ \emph {et~al.}(1994)\citenamefont {Sun},
  \citenamefont {Vall\'ee}, \citenamefont {Acioli}, \citenamefont {Ippen},\
  and\ \citenamefont {Fujimoto}}]{Sun1994}%
  \BibitemOpen
  \bibfield  {author} {\bibinfo {author} {\bibfnamefont {C.-K.}\ \bibnamefont
  {Sun}}, \bibinfo {author} {\bibfnamefont {F.}~\bibnamefont {Vall\'ee}},
  \bibinfo {author} {\bibfnamefont {L.~H.}\ \bibnamefont {Acioli}}, \bibinfo
  {author} {\bibfnamefont {E.~P.}\ \bibnamefont {Ippen}}, \ and\ \bibinfo
  {author} {\bibfnamefont {J.~G.}\ \bibnamefont {Fujimoto}},\ }\href {\doibase
  10.1103/PhysRevB.50.15337} {\bibfield  {journal} {\bibinfo  {journal} {Phys.
  Rev. B}\ }\textbf {\bibinfo {volume} {50}},\ \bibinfo {pages} {15337}
  (\bibinfo {year} {1994})}\BibitemShut {NoStop}%
\bibitem [{\citenamefont {Mazin}\ \emph {et~al.}(2013)\citenamefont {Mazin},
  \citenamefont {Manni}, \citenamefont {Foyevtsova}, \citenamefont {Jeschke},
  \citenamefont {Gegenwart},\ and\ \citenamefont {Valent\'\i}}]{Mazin_13}%
  \BibitemOpen
  \bibfield  {author} {\bibinfo {author} {\bibfnamefont {I.~I.}\ \bibnamefont
  {Mazin}}, \bibinfo {author} {\bibfnamefont {S.}~\bibnamefont {Manni}},
  \bibinfo {author} {\bibfnamefont {K.}~\bibnamefont {Foyevtsova}}, \bibinfo
  {author} {\bibfnamefont {H.~O.}\ \bibnamefont {Jeschke}}, \bibinfo {author}
  {\bibfnamefont {P.}~\bibnamefont {Gegenwart}}, \ and\ \bibinfo {author}
  {\bibfnamefont {R.}~\bibnamefont {Valent\'\i}},\ }\href@noop {} {\bibfield
  {journal} {\bibinfo  {journal} {Phys. Rev. B}\ }\textbf {\bibinfo {volume}
  {88}},\ \bibinfo {pages} {035115} (\bibinfo {year} {2013})}\BibitemShut
  {NoStop}%
\end{thebibliography}%

\newpage

\onecolumngrid
\setcounter{figure}{0}
\renewcommand{\thefigure}{S\arabic{figure}}

\part*{Supplementary Materials}

\begin{figure}[b]
\noindent \begin{centering}
\begin{tabular}{cc}
\includegraphics[height=0.18\textheight]{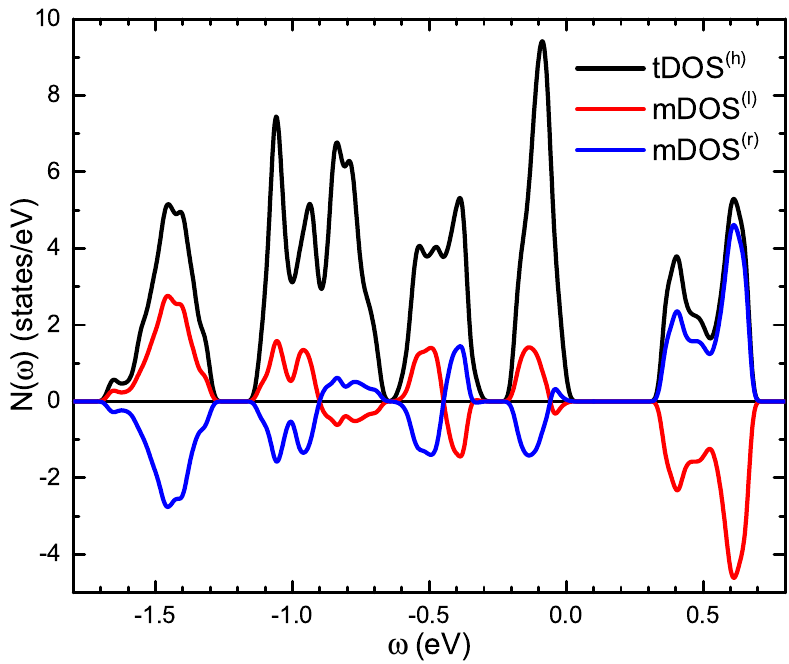} & \includegraphics[height=0.18\textheight]{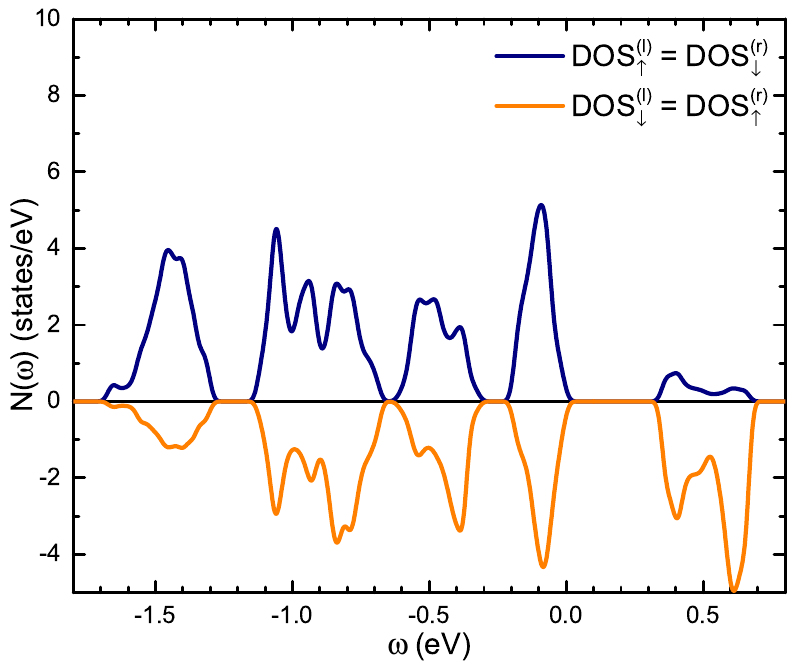}\tabularnewline
\includegraphics[height=0.18\textheight]{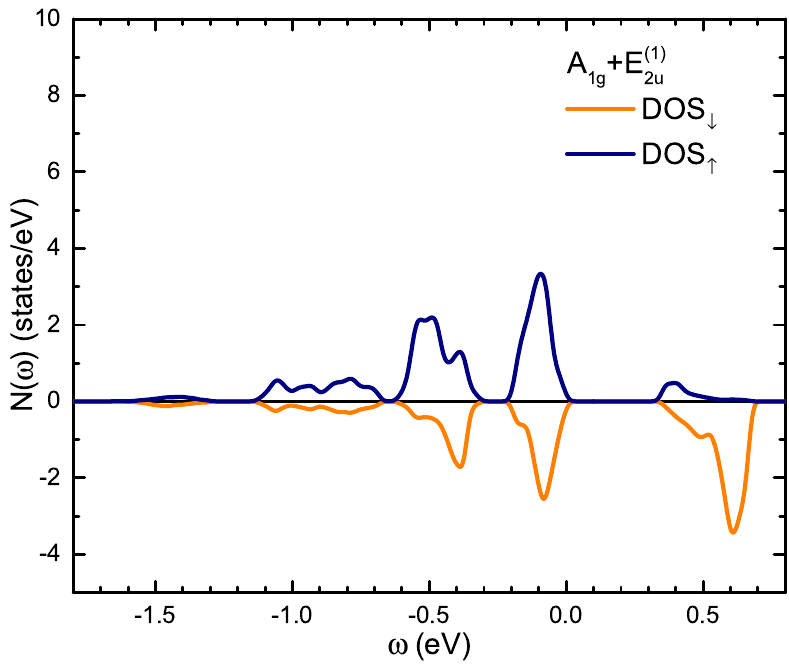} & \includegraphics[height=0.18\textheight]{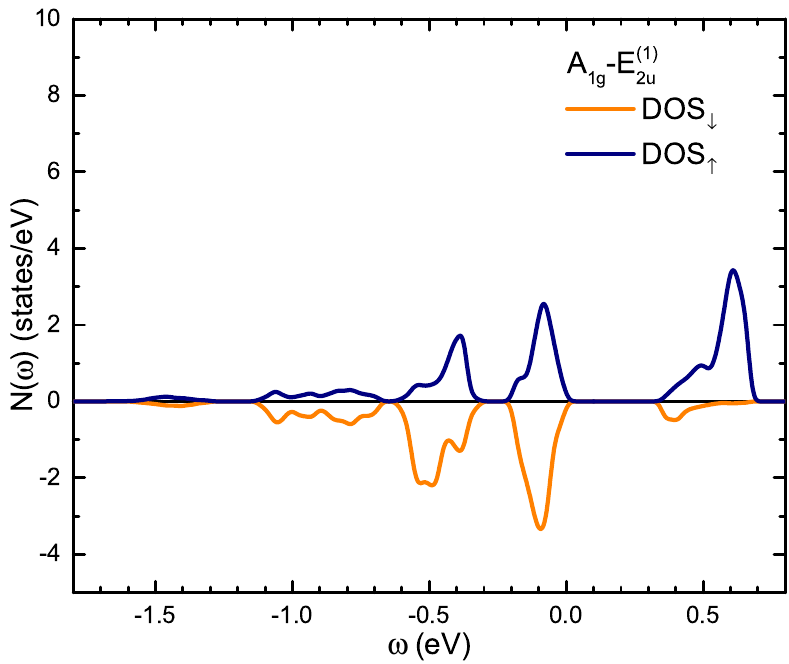}\tabularnewline
\includegraphics[height=0.18\textheight]{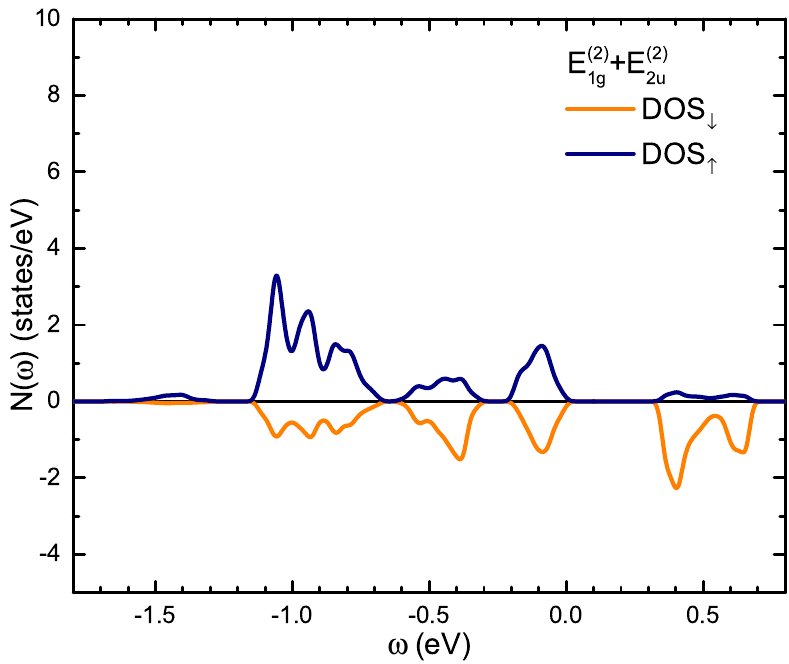} & \includegraphics[height=0.18\textheight]{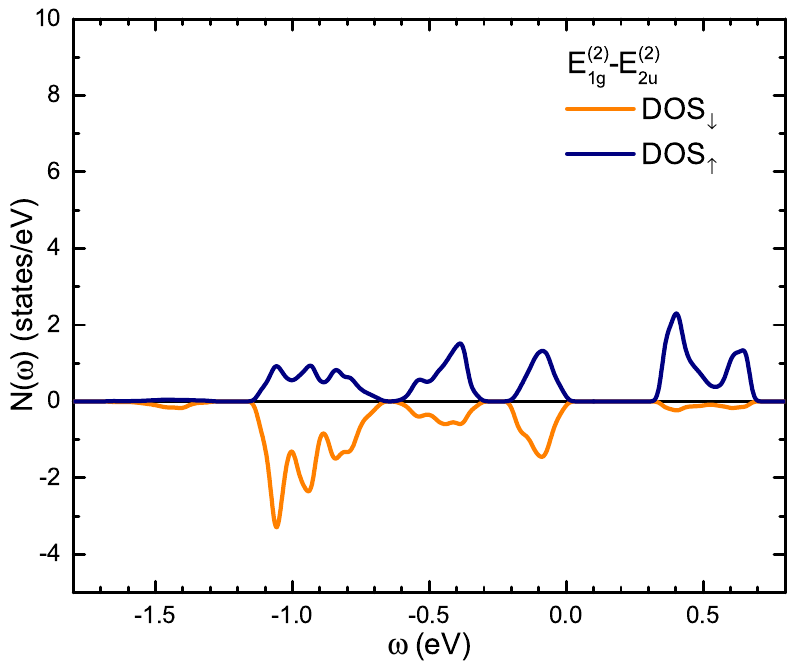}\tabularnewline
\includegraphics[height=0.18\textheight]{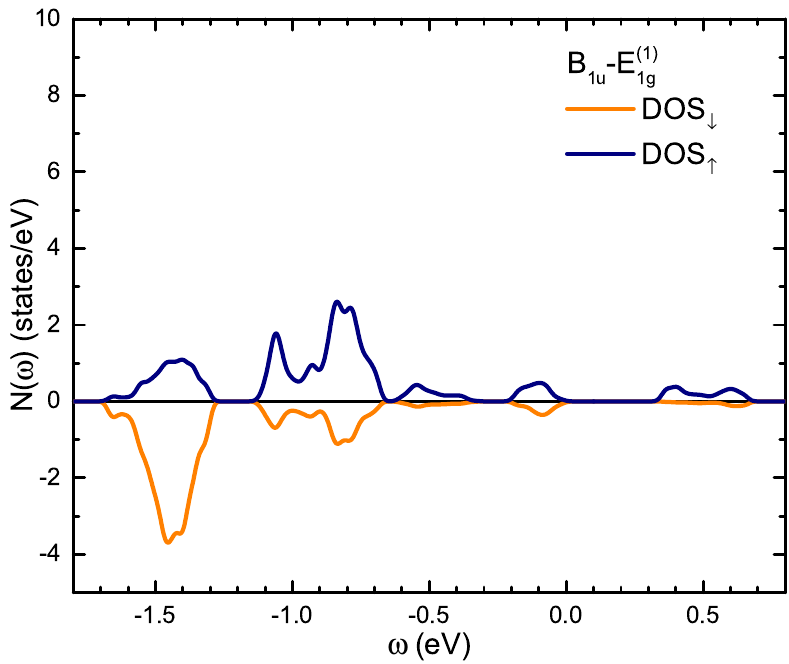} & \includegraphics[height=0.18\textheight]{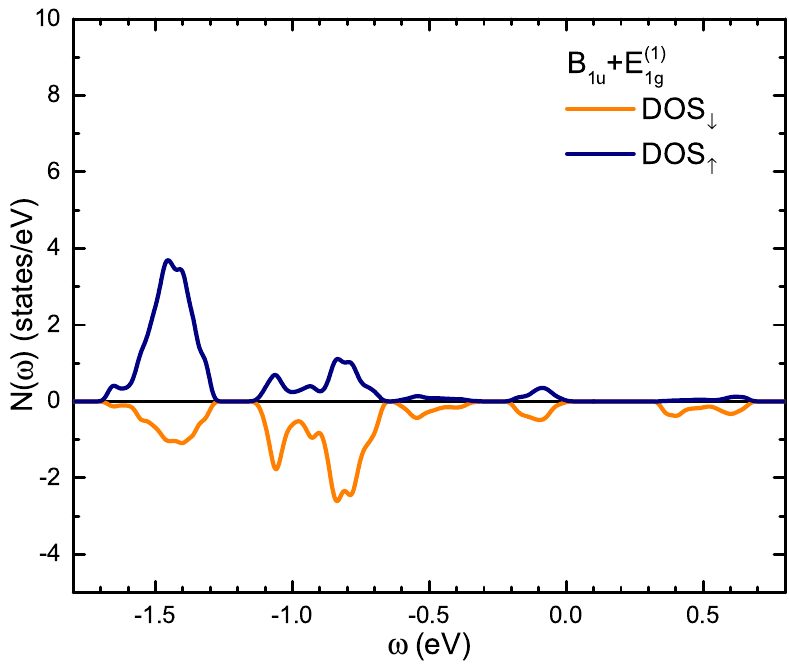}\tabularnewline
\end{tabular}
\par\end{centering}
\caption{DOS calculated by DFT. (a) Total (t) density of states per spin on
one hexagon (h): tDOS$^{\left(h\right)}$ = $\nicefrac{1}{2}$(DOS$_{\uparrow}^{\left(h\right)}$-DOS$_{\downarrow}^{\left(h\right)}$),
net polarization (m) in the left (l) and right (r) half of the hexagon
mDOS$^{\left(l,r\right)}$ = DOS$_{\uparrow}^{\left(l,r\right)}$-DOS$_{\downarrow}^{\left(l,r\right)}$.
(b) Spin-up and spin-down DOS on the left and right half of the hexagon
DOS$_{\uparrow,\downarrow}^{\left(l,r\right)}$. (c-h) Spin-up and
spin-down DOS of the six combinations of QMOs defined in the main
text.\label{fig:DOS-DFT}}
\end{figure}

\section{DFT calculations}

In Fig.~\ref{fig:DOS-DFT}, we report the details of the density
of states (DOS) calculated by DFT+U+SO. Panel (a) reports the total (t) DOS per spin on
one hexagon (h) {[}tDOS$^{\left(h\right)}$ = $\nicefrac{1}{2}$(DOS$_{\uparrow}^{\left(h\right)}$-DOS$_{\downarrow}^{\left(h\right)}$){]}
together with the net polarization (m) in the left (l) and right (r)
half of the hexagon {[}mDOS$^{\left(l,r\right)}$ = DOS$_{\uparrow}^{\left(l,r\right)}$-DOS$_{\downarrow}^{\left(l,r\right)}${]}.
In the tDOS$^{\left(h\right)}$, the five structures (one quite large
centered at $\approx\unit[-0.8]{eV}$ coming from the overlap of two
smaller structures) are reminiscent of the six original QMOs. It is
also evident the full polarization of the empty states and the more
diffused (in energy) polarization of the occupied states per half
of the hexagon suggesting a spatial arrangement of the magnetic moments
fully compatible with the zig-zag order. For the sake of completeness,
in panel (b), we also report the spin-up and spin-down components
of the DOS on the left and right half of the same hexagon {[}DOS$_{\uparrow,\downarrow}^{\left(l,r\right)}${]}
that lead to the net polarizations in panel (a). In panels (c-h),
we report the spin-up and spin-down components of the DOS of the six
combinations of QMOs defined in the main text in order to better estimate
the regions in energy where they overlap, independently of the value
of their net polarizations (reported in Fig.~1 of the main text).

By matching the DOS reported in Fig. \ref{fig:DOS-DFT} to the polarization shown in Figure 3 of the main text, it is possible to link the ground state zig-zag pattern to the combinations of QMOS described above.
The $B_{1u}\pm E_{1g}^{(2)}$ modes are mostly confined to binding
energies of the order of $\unit[0.5-1.5]{eV}$, with a resulting
very small overlap with the states at the Fermi level (Fig. \ref{fig:DOS-DFT}),
and feature a quite small overall net polarization (DOS$\uparrow$-DOS$\downarrow$
shown in Fig. 3 of the main text). Instead, the $E_{1g}^{(1)}\pm E_{2u}^{(2)}$
modes, which are mainly located at $\approx\unit[1]{eV}$ binding energy,
are almost fully polarized according to the zig-zag pattern (dark
red curves in Fig. 3). Furthermore, they are characterized by a non-zero
overlap with both the occupied states right below the Fermi level
and the empty states at $\approx\unit[+0.3]{eV}$ (Fig. \ref{fig:DOS-DFT}), thus constituting a possible bridge between high-energy QMOs and the local magnetic dynamics that involve the low-energy degrees of freedom. However, when we move closer to the Fermi level, the full DOS is recovered only by considering also the contribution of the $A_{1g}\pm E_{2u}^{(1)}$
modes, which are fully polarized (dark green curves in Fig. 3)
and extend in energy from $\approx\unit[-0.3]{eV}$ to $\approx\unit[+0.7]{eV}$
(see Fig. \ref{fig:DOS-DFT}). The necessity of including more than one QMO combination ($E_{1g}^{(1)}\pm E_{2u}^{(2)}$ and $A_{1g}\pm E_{2u}^{(1)}$)
to describe the valence (LHB, in the Hubbard description) and the
conduction (UHB) bands, suggests that the QMO picture, which captures
the main features of the deep electronic states, is less efficient
to describe the physics of the low-energy electronic excitations.
\begin{figure}[t]
\noindent \centering{}\includegraphics[width=1\textwidth]{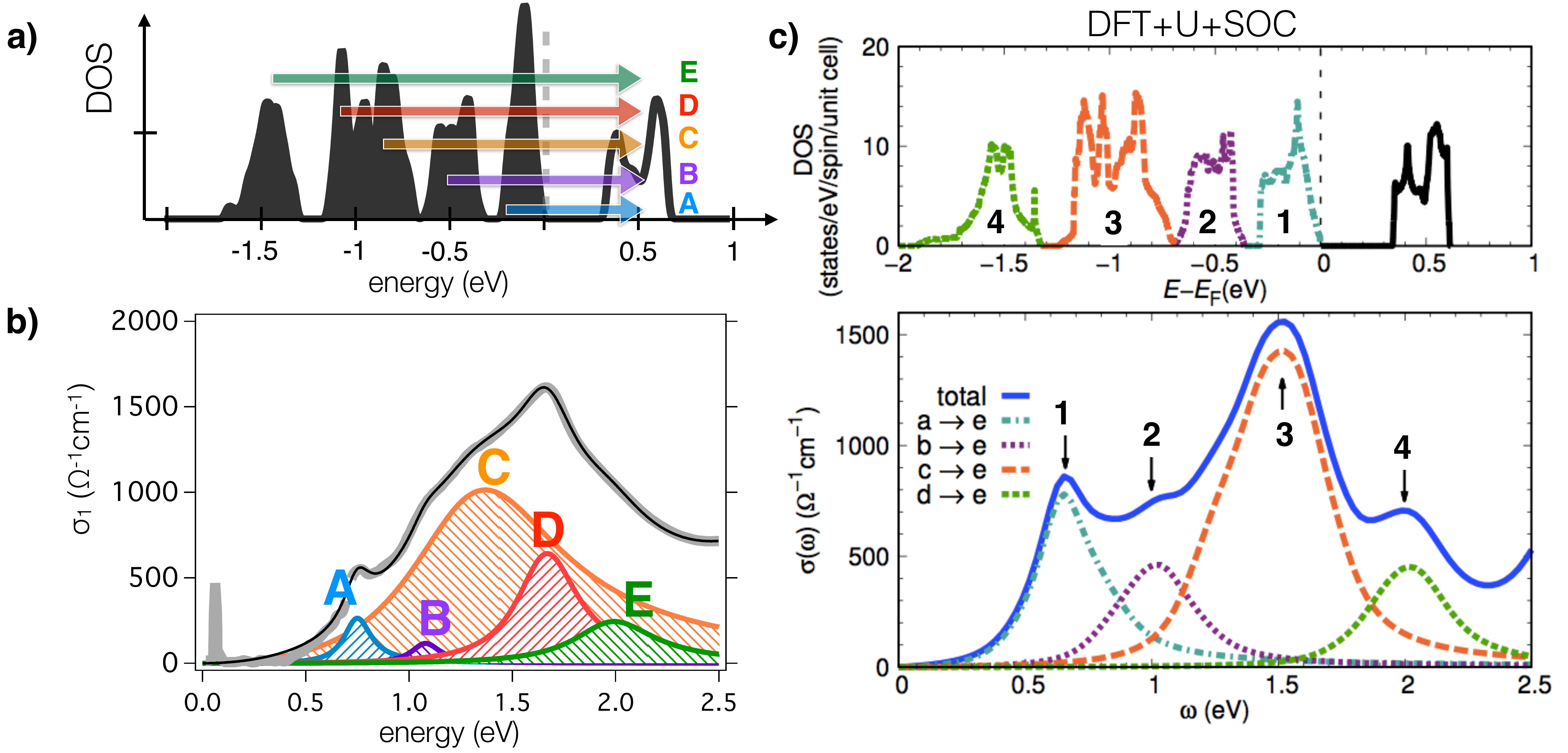}
\caption{a) DFT+U+SO calculated total DOS (see main text). The coloured arrows indicate the possible transitions from the occupied bands to unoccupied levels. b) The real part of the optical conductivity (in practical units, i.e. $\sigma_1\cdot\pi/15$) of Na$_2$IrO$_3$ at 300 K (grey line, from Ref. \citenum{Comin_12}) is plotted as a function of the photon energy. The coloured areas indicate the contribution of the different Lorentz oscillators that are determined by fitting (black line) a multi-peak model to the experimental conductivity. Tha parameters used in the fit are reported in Table I. c) Optical conductivity calculated  within DFT with the same parameters used in the calculations reported in the main text. Taken from Ref. \citenum{Li_15}. \label{fig_S2}}
\end{figure}

\section{dielectric function and differential fitting}
Figure \ref{fig_S2}b reports the 300 K optical conductivity (grey line, from Ref. \citenum{Comin_12}) of Na$_2$IrO$_3$. The $\sigma_1(\omega)$ function is reproduced (black line) by a multi-Lorentzian fit. The number of oscillators that optimizes the fit is five, labeled with capital letters from A to E. The output parameters of the fitting procedure are reported in Table I. The five oscillators correspond to the possible optical transitions involving the DFT-calculated QMOs (see top panel of Fig. \ref{fig_S2}). 
For sake of comparison, Fig. \ref{fig_S2}b) reports the optical conductivity calculated within DFT in Ref. \citenum{Li_15}. We stress the one-to-one correspondence of the four structures (labeled 1-4) present in the calculated optical conductivity and the Lorentz oscillators used in the dielectric function model shown in Fig. \ref{fig_S2}a. The intensity of the transitions (1-4 in the calculated $\sigma(\omega)$, A-E in the experimental $\sigma(\omega)$) can be rationalized on the basis of symmetry arguments involving the parities of the initial and final QMOs involved in the optical transitions \cite{Li_15}. The asymmetric peak 3 in the calculated optical conductivity (Fig. \ref{fig_S2}c) is accounted for by two different Lorentz oscillators (C and D, Fig. \ref{fig_S2}b). These structures are reminiscent of the spin-orbit induced lift of degeneracy of the $E_{1g}$ QMOs and can be microscopically explained by projecting the total density of states onto the $E_{1g}^{(1)}\pm E_{2u}^{(2)}$ and $B_{1u}\pm E_{1g}$ combinations, as showed in Fig. 3 of the main text. While the C  transition involves initial states of symmetry $B_{1u}\pm E_{1g}$ that are polarized oppositely to the zig-zag magnetic ground state, the D oscillator accounts for transitions involving the $E_{1g}^{(1)}\pm E_{2u}^{(2)}$ combination that is fully polarized according to the zig-zag pattern. As argued in the main text, the photo-induced perturbation of the zig-zag magnetic order is expected to particularly affect the $E_{1g}^{(1)}\pm E_{2u}^{(2)}$ QMO combination that exhibits the same spatial pattern and spin-polarization of the magnetic ground state.
As a natural consequence, the differential reflectivity shown in Fig. 3 of the main text can be very well reproduced by simply assuming a photo-induced modification of the strength and position of the oscillator D.

\begin{table}\label{table}
\centering
\begin{tabular}{l|c|c|c}

\multicolumn{4}{c}
{\textbf{Na$_2$IrO$_3$ static optical parameters T$=50$ K }}\\
\hline
& Position (eV) & Plasma Frequency (eV) & Width (eV)\\ \hline
A  & 0.75 &0.46 &0.14\\ \hline
B  & 1.08 & 0.36 & 0.17 \\ \hline
C  & 1.39 & 2.88& 0.99 \\ \hline
D  & \textbf{1.65} & \textbf{1.097} & 0.32 \\ \hline
E  & 1.97 & 0.77 &0.46 \\ \hline
\end{tabular}
\caption{Multi-Lorentzian model parameters used in the equilibrium and differential fitting. In black we have highlight the oscillator parameters that should be changed  to obtain the differential fit reported in Fig. 3 of the main text.}
\end{table}

\section{impulsive heating}
\begin{figure}[t]
\noindent \centering{}\includegraphics[width=1\textwidth]{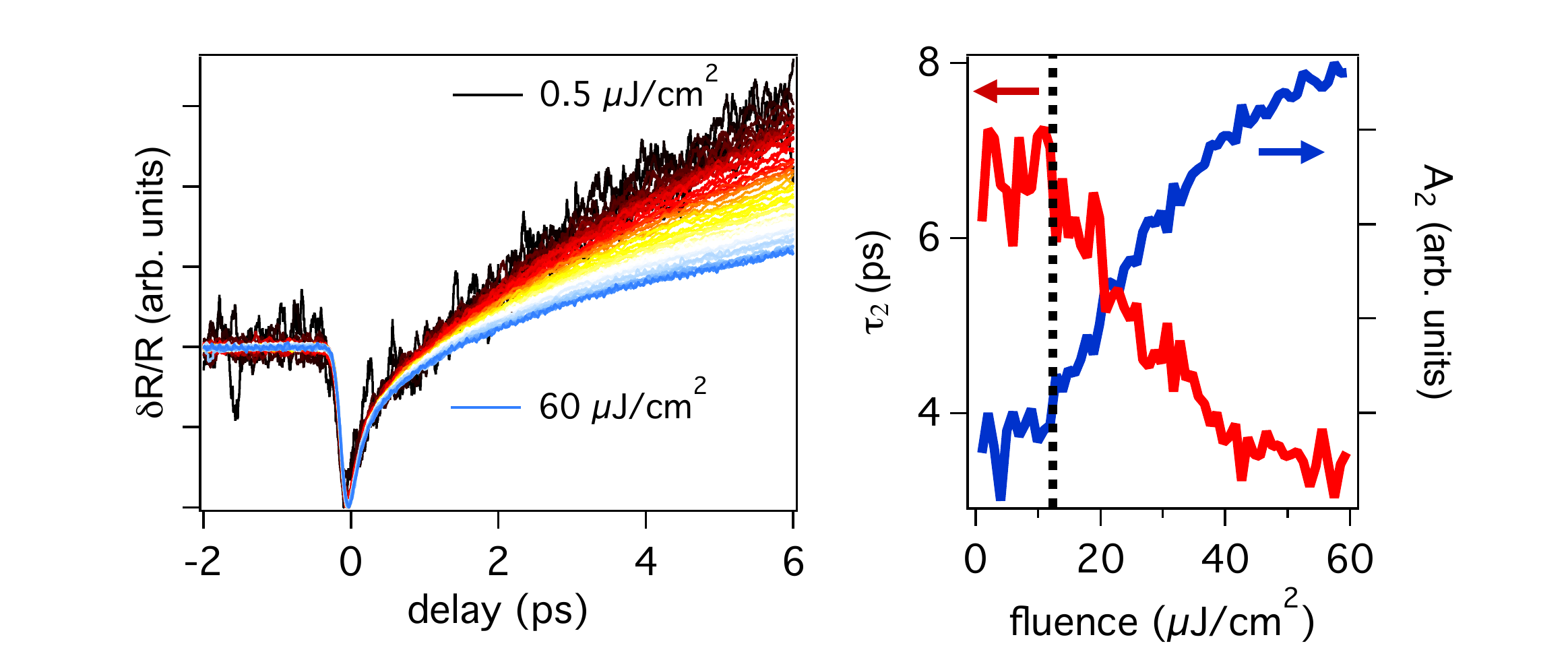}
\caption{Left panel: single-color (1.55 eV) pump-probe measurements at different fluences (0.5-60 $\mu$J/cm$^2$). Right panel: amplitude (A$_2$) and decay time ($\tau_2$) of the slow component as a function of the pump fluence. The values of $\tau_2$ and A$_2$ were extracted by the multiexponential fit (see main text) to the data shown in the left panel.  \label{fig_heating}}
\end{figure}

The laser induced local heating of the sample is an important issue that should be addressed to properly interpret the low-temperature dynamics. While the average heating was controlled by tuning the repetition rate of the Ti:sapphire cavity, the impulsive heating induced by the single pulse was empirically addressed by performing single-color ($\hbar \omega$=1.55 eV) pump-probe measurements in a very broad excitation range. In the left panel of Fig. \ref{fig_heating}, we report the $\delta R/R$(t) signal, normalized to the amplitude of the fast component, at different excitation fluences (0.5-60 $\mu$J/cm$^2$). For each pump fluence, the amplitude (A$_2$) and decay dynamics ($\tau_2$) of the slow component is obtained by performing the two-exponential fitting described in the main text. The values of $\tau_2$ and A$_2$ are constant up to a threshold fluence of about 12 $\mu$J/cm$^2$. Above this value, the A$_2$ signal reduces suggesting a progressive onset of impulsive thermal effects. 

The temperature-dependence of the A$_2$ and $\tau_2$ values reported in Fig. 2 of the main text, has been obtained by fitting single-color pump-probe measurements at a pump fluence of 1 $\mu$J/cm$^2$. Considering a penetration depth of $\sim$ 100 nm and a heat capacity of 74 mJ/cm$^3$K \cite{Singh_10}, we estimate a negligible impulsive heating of the order of 1 K at a base temperature of 20 K. The estimated impulsive heating increases up to about 3 K at temperatures as low as 10 K. 
Interestingly, the experimental finding that the A$_2$ and $\tau_2$ values are constant up to fluences as large as 12 $\mu$J/cm$^2$, where the impulsive heating is of the order of 10 K, further supports the main claim of the manuscript, i.e., that the A$_2$ component is related to the energy exchange with short-range zig-zag correlations on a length scale of 2 nm, that are more robust than the long-range 3D order.

\section{sub-\lowercase{ps} dynamics}
\begin{figure}[t]
\noindent \centering{}\includegraphics[width=0.5\textwidth]{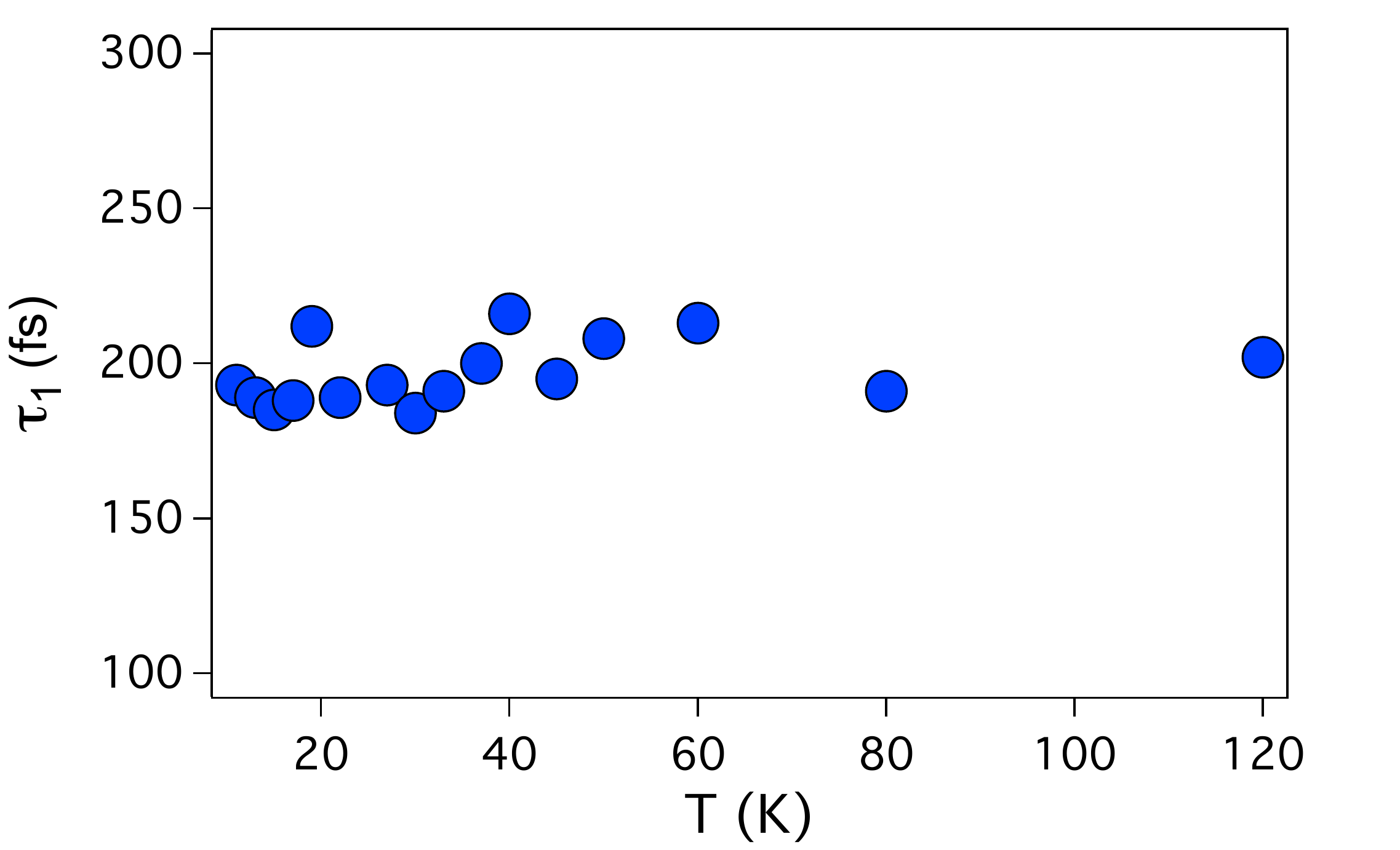}
\caption{The temperature dependence of the fast exponential decay ($\tau_1$) is reported. The values of $\tau_1$ (blue dots) are obtained by the multi-exponential fitting of the time-resolved data reported in Figs. 2,3 of the main text. \label{fig_S3}}
\end{figure}

Figure \ref{fig_S3} reports the temperature-dependent value of the $\tau_1$ time constant, as determined from the multi-exponential time-domain fitting described in the main text. The fast relaxation dynamics has an average value of $\tau_1\approx 200$ fs and it does not display any significant variation when the magnetic transition at $T_N$ is approached. 

\end{document}